%% file: paper.tex
\pdfoutput=1

\documentclass[hyper]{JHEP3}

\usepackage{amsmath}
\usepackage{amssymb}
\usepackage{graphicx}

\newcommand{\bZ}{\mathbb{Z}}
\newcommand{\bC}{\mathbb{C}}
\newcommand{\bI}{\mathbb{I}}
\newcommand{\bF}{\mathbb{F}}
\newcommand{\bR}{\mathbb{R}}
\newcommand{\bP}{\mathbb{P}}
\newcommand{\cN}{\mathcal{N}}
\newcommand{\cM}{\mathcal{M}}

\newcommand{\Tr}{\mathrm{Tr\,}}
\newcommand{\ov}{\overline}

\newcommand{\vev}[1]{\left\langle#1\right\rangle}

\newcommand{\color}[2][]{}

\newcommand{\drawsquare}[2]{\hbox{%
\protect\rule{#2pt}{#1pt}\protect\hskip-#2pt%  left vertical
\protect\rule{#1pt}{#2pt}\protect\hskip-#1pt%  lower horizontal
\protect\rule[#1pt]{#1pt}{#2pt}}\protect\rule[#1pt]{#2pt}{#2pt}\protect\hskip-#2pt%  upper horizontal
\protect\rule{#2pt}{#1pt}}% right vertical

\newcommand{\praisebox}[1]{\protect\raisebox{#1}}

\newcommand{\fund}{\praisebox{-.5pt}{\drawsquare{6.5}{0.4}}}% fund
\newcommand{\antifund}{\overline{\fund}}

\title{Toric Lego: A method for modular model building}

\author{Vijay Balasubramanian$^\clubsuit$, Per Berglund$^\spadesuit$$^\heartsuit$
  and I\~naki Garc\'ia-Etxebarria$^\clubsuit$\\
  $\clubsuit$ Department of Physics and Astronomy, University of Pennsylvania,\\
  Philadelphia, PA 19104-6396, USA\\
  $\spadesuit$ Department of Physics, University of New Hampshire,\\
  Durham, NH 03824, USA\\
   $\heartsuit$ PH-TH Division, CERN, CH-1211 Geneva 23, Switzerland\\
  E-mail: \email{vijay@physics.upenn.edu},
  \email{per.berglund@unh.edu}, \email{inaki@sas.upenn.edu}}

\abstract{Within the context of local type IIB models arising from
  branes at toric Calabi-Yau singularities, we present a systematic
  way of joining any number of desired sectors into a consistent
  theory. The different sectors interact via massive messengers with
  masses controlled by tunable parameters. We apply this method to a
  toy model of the minimal supersymmetric standard model (MSSM)
  interacting via gauge mediation with a metastable supersymmetry
  breaking sector and an interacting dark matter sector. We discuss
  how a mirror procedure can be applied in the type IIA case, allowing
  us to join certain intersecting brane configurations through massive
  mediators.}

\preprint{CERN-PH-TH/2009-197\\
UNH-09-04\\
UPR-1213-T}

\begin{document}

\section{Introduction and methodology}

Many interesting and successful classes of models for physics beyond
the standard model involve various different sectors of light modes
coupled only through massive mediator particles. For example, models
of gauge mediated supersymmetry breaking (GMSB) are composed of a
sector representing the degrees of freedom in the MSSM, and another
sector breaking supersymmetry (for a review,
see~\cite{Giudice:1998bp}). These sectors are coupled by massive
mediators which induce soft supersymmetry breaking terms in the
visible sector once we integrate out the massive fields.  Recent
observations by the PAMELA, ATIC and FERMI collaborations have also
suggested the possibility of an additional interacting dark mater
sector, coupled to the standard model via massive mediators
\cite{ArkaniHamed:2008qn}.

It is therefore interesting to devise methods for embedding such
multi-sector interacting models in string theory.  From the
perspective of the landscape of string theory vacua there are also
statistical reasons to expect multiple interacting sectors - all we
really require in a model is that below a TeV we should recover the
standard model.  Previous work has embedded two-sector models of gauge
mediated supersymmetry breaking in string theory
\cite{Diaconescu:2005pc,GarciaEtxebarria:2006aq,GarciaEtxebarria:2006rw}.\footnote{Alternative
  mechanisms for mediation of supersymmetry breaking which have been
  engineered in string theory include (higher form) $U(1)$ mediation
  \cite{Verlinde:2007qk,Grimm:2008ed}, anomaly mediation
  \cite{Rattazzi:2003rj,Choi:2005ge,Kachru:2007xp}, instanton
  mediation \cite{Buican:2008qe}, holographic mediation
  \cite{Benini:2009ff} and of course gravity mediation
  \cite{Kaplunovsky:1993rd,Brignole:1993dj}.} Here we extend these
ideas to a systematic procedure for joining theories coming from
branes at toric singularities into a consistent local embedding, such
that the different subsectors interact only through massive mediators,
with tunable scales.  This allows us to construct string theoretic
embeddings of theories with an arbitrary number of hidden sectors,
such as those advocated in models of interacting dark matter
\cite{ArkaniHamed:2008qn}.  From the field theoretic perspective, we
are giving a systematic procedure for adding massive mediators between
subsectors, such that the whole model is embeddable in string theory,
with mediator masses geometrized as the sizes of resolved cycles in a
singularity.  Our procedure allows us to join subsectors in a modular
way, and covers a large class of interesting models, such as branes at
abelian orbifolds, or branes wrapping local cones over toric del Pezzo
surfaces.

Our framework is a refinement of
\cite{GarciaEtxebarria:2006aq,GarciaEtxebarria:2006rw} where it was
shown how local singular toric Calabi-Yau geometries can be built from
a slightly resolved parent singularity in order to construct
multi-sector models interacting through massive matter in Type II
string theory.  These papers did not give a recipe for finding the
particular parent theory that can be resolved to give two subsectors
of choice.  This was because the same geometry can support different
field theories related by Seiberg duality, which is realized as toric duality in the 
geometry~\cite{Seiberg:1994pq,
  Feng:2000mi}.
When building a particular model, one is often interested in getting a
particular representative of the family of Seiberg duals after
resolution -- but this typically involves trial and error with many
Seiberg dual phases.  Instead of trying to intuit the correct parent
singularity, we describe here a bottom-up approach: we give a method
of splicing together sectors that realize the desired low-energy
phenomenology.

The theories we construct have a very rich spectrum of massive
mediators, with a plethora of both chiral and vector
multiplets. Despite their complexity as low energy theories, they are
remarkably simple from the string theory point of view, depending only
on a small number of parameters with simple geometric realizations.
Such rich mediators are generic in models arising from D-branes at
sub-stringy separation.

\medskip

This paper is organized as follows. Section~\ref{sec:review} reviews
some basic facts about branes at toric singularities and the
corresponding dimer models that we need. Section~\ref{sec:joining}
describes our algorithm for joining models and their corresponding
singularities, a key technical result in this
paper. Section~\ref{sec:dark-mssm} describes a particular three-sector
dark matter model.  Section~\ref{sec:other-sectors} shows how the
construction can be modified, including the introduction of
orientifold planes.  Section~\ref{sec:mirror} presents the type IIA mirror
description of joining dimer models, and section~\ref{sec:conclusions}
concludes the paper.

While this paper was being completed we became aware of
\cite{Amariti:2009tu}, which contains results related to ours.

\section{Review of dimer model technology}
\label{sec:review}

We start by briefly reviewing the aspects of dimer models that we
 use in the present work.  The reader familiar with dimer model
techniques can safely jump to section~\ref{sec:joining}, although
section~\ref{sec:partial-resolution} reviews less well known
material. For further details, the interested reader may consult the
excellent reviews by Kennaway \cite{Kennaway:2007tq} and Yamazaki
\cite{Yamazaki:2008bt}.

\subsection{Toric geometry: web diagrams and toric diagrams}
\label{sec:toric-geometry}

Let us start by recalling some elementary facts about toric Calabi-Yau
threefolds. We use the conifold as our example, with other toric
varieties constructed in a similar fashion. In its singular limit,
the conifold is defined by the
equation
\begin{equation}
  \sum_{i=1}^4 z_i^2 = 0
  \label{conifold}
\end{equation}
inside $\bC^4$. This geometry is toric, meaning that it admits a
$U(1)^3=T^3$ fibration, this can be seen as
follows. Equation~\eqref{conifold} is invariant under $SO(4)$
rotations transforming the $z_i$ as a vector, and also invariant under
$U(1)$ rotations of the four coordinates, sending $z_i\to
e^{i\theta}z_i$. This gives a symmetry group $SO(4)\times U(1)\sim
SU(2)\times SU(2) \times U(1) \supset U(1)^3$, so the conifold is
indeed a toric variety.  In our examples, we are interested
in a smooth space connected to the singular conifold described above,
the \emph{resolved conifold}, obtained by blowing up a two sphere of
size $v$ at the singularity of the conifold.

\FIGURE{
  
\ifpdf
  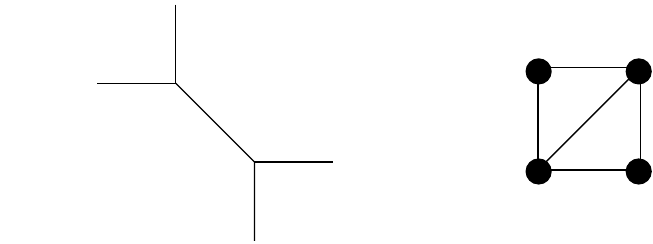
  \else
  \input{coniresolv.pstex_t}
\fi

  \caption{a) Web diagram for the conifold. We have mapped the
    non-compact divisors of the conifold to regions in the web
    diagram. $v$ denotes the resolution parameter. b) Toric diagram
    for the conifold.}

  \label{fig:coniresolv}
}

It turns out that all the relevant information about the geometry can
be efficiently encoded by specifying over which points in the (three
real dimensional) base the $T^3$ fiber degenerates. For the case of
Calabi-Yau manifolds, this information can be easily visualized in
terms of a \emph{web diagram}.  The web diagram is a projection of the
base of the toric manifold onto a plane. The base is a three real
dimensional polyhedron with a $T^3$ fiber over each point -- the
fibers degerate over faces and edges of the polyhedron which appear in
the planar projection of the polyhedron (see \cite{vafaleung} for a
review).  For our purposes it  suffices to know that every manifold
posses a web diagram, which is a graph drawn on a integer lattice.
The web diagram for the resolved conifold is given in
figure~\ref{fig:coniresolv}. The singular conifold is recovered by
shrinking away the edge marked $v$. We also show its planar dual,
the \emph{toric diagram}, which encodes the same information and is
often more useful to work with.

\subsection{Gauge theories from toric singularities: quiver diagrams and dimer models}

Let us now proceed to put a stack of $N$ D3 branes on the singular
point $z_i=0$. As argued by Klebanov and Witten
\cite{Klebanov:1998hh}, the resulting theory is $U(N)_1\times U(N)_2$
with four bifundamentals\footnote{We  always use the notation
  $X^\bullet_{ab},Y^\bullet_{ab},\ldots$ to denote fields transforming in the
  fundamental representation of $U(N)_a$ and the antifundamental of
  $U(N)_b$.} $X_{12}^i$, $X_{21}^i$ ($i$ takes values in ${1,2}$), and
a superpotential:
\begin{equation}
\label{eq:Wconifold}
  W = \Tr (X_{12}^1 X_{21}^1 X_{12}^2 X_{21}^2 - X_{21}^2 X_{12}^2
  X_{21}^1 X_{12}^1)
\end{equation}

\FIGURE{
  
\ifpdf
  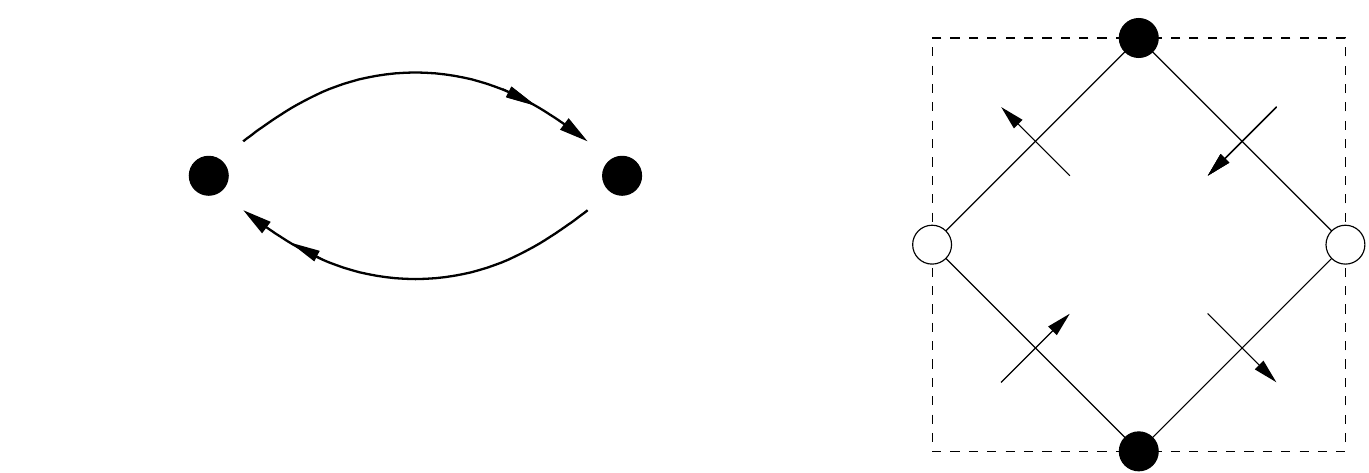
  \else
  \input{coniquiverdimer.pstex_t}
\fi

  \caption{a) Quiver diagram for the conifold. The arrow with two
    heads represents two bifundamentals. b) Dimer model for
    the conifold. Note that the opposite sides of the square should be
    identified, forming a two-torus, i.e., there is a single black
    dot and a single white dot. We have denoted the orientation of
    the edges by arrows perpendicular to them: an arrow from $U(N)_a$
    to $U(N)_b$ crossing the edge $E$ means that $E$ transforms in
    $(\fund_a,\antifund_b)$. We have also labelled the bifundamentals
    as in the text.}

  \label{fig:coniquiverdimer}
}

It is helpful to represent this information graphically. The most
common representation is in terms of \emph{quiver diagrams}. These are
oriented graphs in the plane where nodes represent gauge groups, and
bifundamentals fields are arrows connecting the nodes. We show the
quiver diagram for the conifold in
figure~\ref{fig:coniquiverdimer}a. This representation of the theory,
however, misses the F-term information encoded in the
superpotential. It turns out that for theories coming from D3 branes
located at toric singularities, such as the conifold, it is possible
to do much better, using \emph{dimer models} \cite{Franco:2005rj}. A
dimer model is a periodic tiling of ${\bf R}^2$ (equivalently, a
tiling of $T^2$) with the property that nodes in the graph can be
colored black or white, and white nodes only connect to black nodes
(and vice versa).  The dictionary between this tiling and properties of the associated gauge theory is as follows:
\begin{itemize}
\item Faces represent $U(N)$ gauge groups.\footnote{In string theory
    most of the $U(1)$ factors actually decouple at low energies,
    getting mass by mixing with background $RR$ axions
    \cite{Ibanez:1998qp,Morrison:1998cs}, although some of the
    anomaly-free $U(1)$ factors can remain massless. For simplicity,
    we  always keep these $U(1)$ factors around, and just discuss
    their fate when it makes a difference.\label{U(1)-footnote}} In
  the conformal case we have discussed so far all faces have the same
  rank, but we will see below that it is also possible that faces have
  different rank, representing D5 and D7 brane charge localized at the
  singularity (i.e., fractional D3 charge).
\item Edges between faces $U(N)_a$ and $U(N)_b$ represent
  bifundamental fields. It is possible to assign a unique orientation
  to the edges (whether they transform in $(\fund_a,\antifund_b)$ or
  in $(\antifund_a,\fund_b)$) by imposing that edges around a black
  node go clockwise and around a white node counter-clockwise. Note
  that this implies that as we go around a face, edges alternate in
  their orientation. This also implies that every face has an even
  number of edges, half incoming and half outgoing.
\item Finally, nodes represent superpotential terms (more precisely,
  monomials in the superpotential). The rule to obtain the precise
  monomial is to multiply the fields around a black node clockwise,
  with an overall plus sign, and multiply fields around a white node
  counter-clockwise, with an overall minus sign. One should take the
  trace of the resulting polynomial in order to obtain a gauge
  invariant superpotential.
\end{itemize}
We have shown the quiver diagram and dimer model for the conifold in
figure~\ref{fig:coniquiverdimer}. It is a simple exercise to verify
that with the rules given here the dimer model in
figure~\ref{fig:coniquiverdimer}b encodes precisely the conifold
theory described above. In particular, the superpotential is given by~(\ref{eq:Wconifold}).

At this point two natural questions arise: can we obtain the dimer
diagram for a given toric singularity in a simple way? Is the geometry
of the toric variety naturally encoded in the dimer
diagram?\footnote{The toric variety should
  arise as the moduli space for the theory of a single regular brane
  located at the singular point, so the dimer model does encode all
  the information about the toric singularity. The question is how to
  determine this moduli space efficiently from the dimer model.}
Surprisingly, the answer to both questions is affirmative, the
essential concept being the \emph{zig-zag
  path}~\cite{kenyon-2005-357,Hanany:2005ss}.  We now review the
definition and properties of zig-zag paths, which are an essential
tool in our construction, and show how they naturally give an elegant
answer to the second question. We postpone the answer to the first
question to section~\ref{sec:fast-inverse}.

\subsection{Zig-zag paths: From dimer models to toric geometry}

\FIGURE{
  \includegraphics[width=0.6\textwidth]{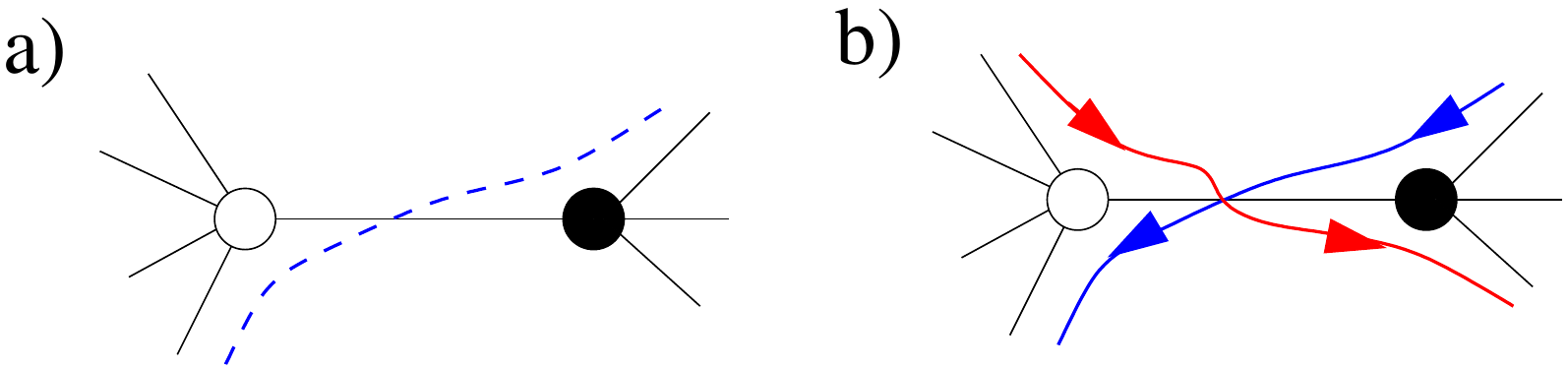}

  \caption{a) Generic zig-zag path, drawn as a dashed blue line. It
    turns maximally when it reaches an edge, and it crosses the edges
    along which it runs. b) When two zig-zag paths intersect, they do
    it over an edge and running with opposite orientations.}

  \label{fig:zzpath}
}

Zig-zag paths \cite{kenyon-2005-357,Hanany:2005ss} (also called
\emph{rhombus paths} in the context of isoradial embeddings) are paths
in the dimer model with the property that they turn maximally at each
node, and they cross once the edges along which they run. We
illustrate a typical zig-zag path in figure~\ref{fig:zzpath}a (see
also figure~\ref{fig:conizz} for the set of zig-zag paths for the
conifold).

Zig-zag paths have a number of interesting and useful properties, of
which we mention a few here. We do not provide justification for the
following statements, but refer the reader to the original papers
\cite{kenyon-2005-357,Hanany:2005ss,Feng:2005gw}.

The first property is that each edge in the dimer model corresponds to
an intersection of two zig-zag paths. That is, every time two zig-zag
paths intersect they give rise to a bifundamental field.

A second property is that we can naturally assign an orientation to
each zig-zag path. In order to do this, assign arbitrarily an
orientation to one of the zig-zag paths. The orientation of the rest
of the zig-zag paths is determined by the requirement that when two
zig-zag paths intersect over an edge, they do it with opposite
orientations, see figure~\ref{fig:zzpath}b.

We can now describe how zig-zag paths solve the second problem above,
namely how to obtain the toric variety from the dimer model.  Take one
zig-zag path, with its orientation. Compute its $(p,q)$ winding number
along the torus on which the dimer model is defined. This can be
represented as a vector in the plane, pointing in the $(p,q)$
direction. Once we do this for all zig-zag paths, we
obtain the web diagram of the (unresolved) toric singularity.\footnote{This can be
  naturally understood using mirror symmetry~\cite{Feng:2005gw}. See
  also section~\ref{sec:mirror}.} As an example, in
figure~\ref{fig:conizz} we compute the web diagram of the conifold
from the conifold dimer model.

\subsection{The fast inverse algorithm: From toric geometry to dimer models}
\label{sec:fast-inverse}

Let us now proceed to describe how one can compute the dimer model
corresponding to a given toric singularity. The most efficient
algorithm, and the one that will inspire our joining algorithm in
section~\ref{sec:joining}, is the \emph{fast inverse algorithm}, first
described in \cite{Hanany:2005ss}. Since we will make use of some of
the ideas of this algorithm in the following sections, let us briefly
describe how it works.

\FIGURE{
  \includegraphics[width=0.8\textwidth]{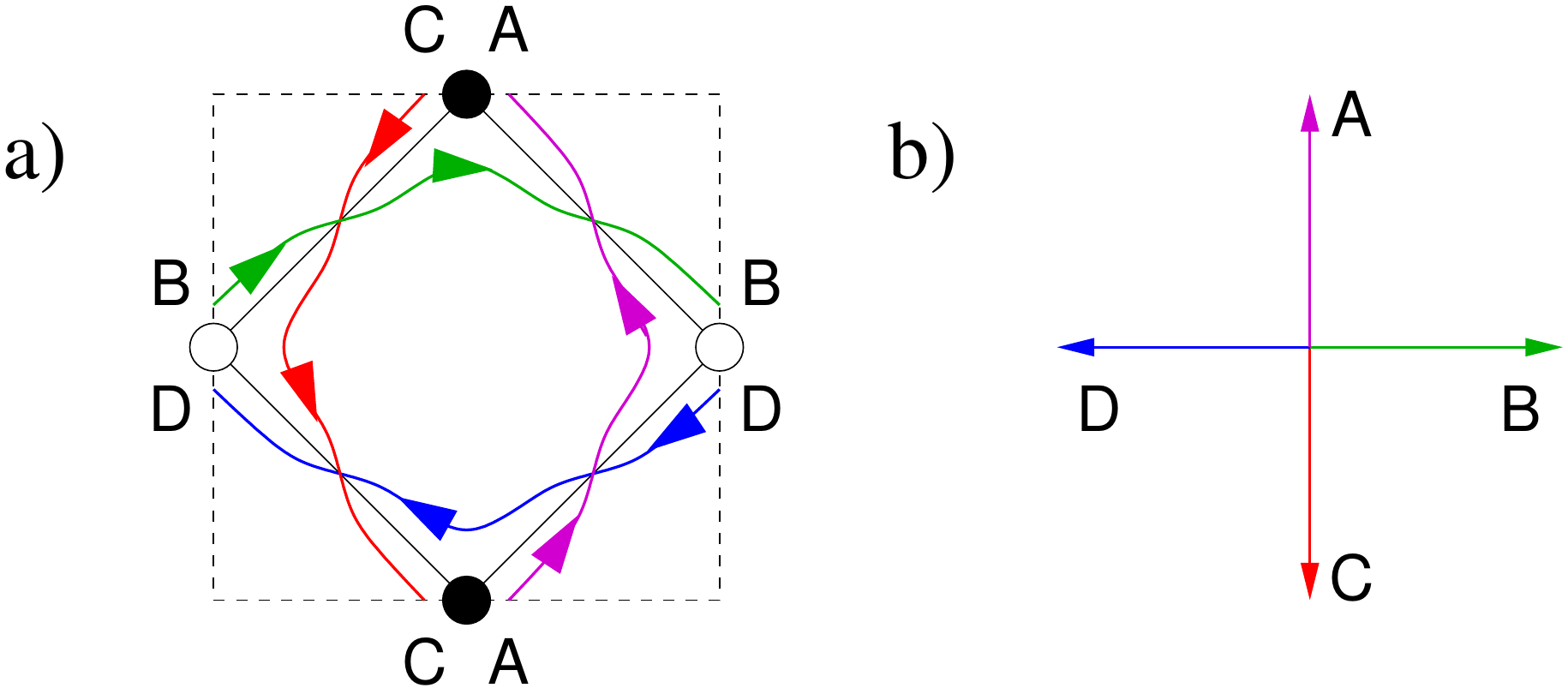}

  \caption{a) Set of zig-zag paths for the conifold. b) The web diagram
    obtained out of the zig-zag paths.}

  \label{fig:conizz}
}

The idea is simple and uses the key concept of zig-zag paths: draw in a
torus cycles with homology charges corresponding to the external legs
of the geometry we want to consider. These cycles divide the torus
into regions, with boundaries given by portions of zig-zag
paths. These regions can be divided into two classes: those around
which the boundary has a definite orientation (inherited from the
zig-zag paths defining the boundary), and those which have no definite
orientation. From this information we can easily recover the 
dimer model, simply by identifying oriented regions with
superpotential terms, zig-zag path intersections with edges, and
unoriented regions with gauge factors. We show how this works in the
case of the conifold in figure~\ref{fig:coni-inverse}. Let us note
that we could have further subdivided oriented regions into clockwise
and counter-clockwise regions. As it is simple to see from examples,
this  corresponds to the black/white coloring of the nodes in the
dimer model.

\FIGURE{
  \includegraphics[width=0.4\textwidth]{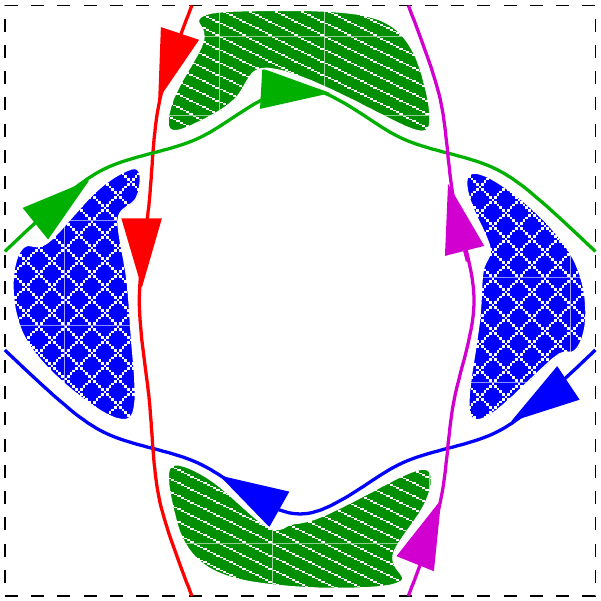}

  \caption{Inverse algorithm for the conifold. The two shaded regions
    are oriented, and correspond to the two superpotential terms,
    while the two regions not shaded do not have a definite
    orientation, and correspond to the two faces of the dimer
    model. Intersections of zig-zag paths correspond to bifundamental
    fields.}

  \label{fig:coni-inverse}
}

\subsection{The low energy spectrum for partial resolution}
\label{sec:partial-resolution}

Zig-zag paths have another very interesting application. As mentioned
above in section~\ref{sec:toric-geometry}, it is possible to smooth
out the singularity in the conifold by blowing up a
two-cycle. Similarly, any toric Calabi-Yau variety can be completely
smoothed out by blowing up two and four cycles at the singularity.  A
complete resolution results in a toric diagram which is maximally
triangulated, and alternate triangulations describe resolutions
related by flops.  It is always possible to completely smooth out a
toric Calabi-Yau in this way, but it is not necessary to completely
smooth it out, \emph{partial} resolutions are also possible. In such a
case, one blows up one of the zero-size two-cycles, while leaving
parts of the geometry singular. Each such partial resolution
partitions the toric diagram into concatenated convex polytopes.

As a simple two complex dimensional example that illustrates the idea
of partial resolution, consider the geometry defined by the equation:
\begin{equation}
  f = x^2 + y^2 + z^4 = 0
\end{equation}
in $\bC^3$. This geometry has a singularity (of type $A_3$) at the
origin, due to the fact the $f=df=0$ there. It is possible to blow up
a two-cycle and make the geometry less singular by modifying the
defining equation to:
\begin{align}
  \widetilde f = x^2 + y^2 + z^2 (z-a)^2 = 0
\end{align}
with $a\neq 0$. The new geometry has two singularities of lower rank
(they are now of type $A_1$) at $x=y=z=0$ and at $x=y=0,z=a$. The
toric threefold examples we construct behave in a similar way. We
start with a ``big'' singularity, and partially blow it up in order to
obtain a Calabi-Yau variety with interesting ``smaller'' singularities
at a finite distance from each other.

In our constructions the gauge theory comes from branes located at the
singularity, so we need to know how the geometric operation of
resolving the singularity affects the gauge theory living on the branes. This
problem was solved for arbitrary resolutions of toric Calabi-Yau
varieties in
\cite{GarciaEtxebarria:2006aq,GarciaEtxebarria:2006rw}. As in the rest
of this review section, we merely present the rules for obtaining
the answers, leaving the justification to the original references.

Let us review the algorithm for obtaining two dimer models
corresponding to the massless matter living on the branes at the two
remaining daughter singularities, and the massive mediators coupling
the two massless sectors. In order to fix notation, let us call the
two daughter singularities 1 and 2. As noted above, we can naturally
split the external legs of the toric diagram into two sets, associated
to the two daughter singularities. This naturally gives us a splitting
of the zig-zag paths for the mother singularity into two sets. Let us
abuse notation and also call these sets 1 and 2 (which particular
meaning of ``1'' and ``2'' we are talking about will be clear from the
context).

Having split the zig-zag paths into two sets, associated to the two
daughter singularities, we can classify the set of edges in the parent dimer
into three sets: those over which two zig-zag paths of type 1
intersect (edges of type 1), those over which zig-zag paths of type 2
intersect (edges of type 2), and edges over which a type 1 zig-zag
path and a type 2 zig-zag path intersect (edges of type 3).

The result in \cite{GarciaEtxebarria:2006aq,GarciaEtxebarria:2006rw}
states that the correct field theory description of the resolution of
the singularity is to give the following sets of vevs to the different
types of bifundamentals:
\begin{align}
  \vev{X^1} = \begin{pmatrix}
    0 & 0 \\ 0 & v\bI_Q
    \end{pmatrix}, \qquad
    \vev{X^2} = \begin{pmatrix}
      v\bI_P & 0 \\ 0 & 0
    \end{pmatrix}, \qquad
    \vev{X^3} = 0
\end{align}
where we have denoted edges of type $i$ as $X^i$, and we have split
the $N$ branes at the original singularity into $P$ branes going to
the daughter singularity 1, and $Q$ going to the daughter singularity
2. $\bI_P,\bI_Q$ denote the identity matrices of rank $P,Q$. $v$
parameterizes the size of the blown-up cycle. For simplicity, we have
presented the case in which all the faces of the original dimer have
the same rank $N$ (corresponding to the case with vanishing fractional
brane charge), so the vevs for the $X^i$ bifundamentals can be
naturally represented as $N\times N$ matrices. The case with
non-vanishing fractional charge works similarly (see \cite{GarciaEtxebarria:2006aq,GarciaEtxebarria:2006rw} for details).

This prescription admits a very simple description in terms of dimer
models: edges of type 1 disappear from the dimer model for the
daughter singularity 2 (representing the Higgsing coming from the
nonzero vev), edges of type 2 disappear from the dimer model for the
daughter singularity 1, and edges of type 3 remain in both daughter
dimers. We illustrate this procedure for the case of the conifold in
figure~\ref{fig:coniresolvdimer}.\footnote{In the case of the conifold
  a single blow-up completely smooths out the space, giving two copies
  of flat space, so the term ``daughter singularity'' is perhaps a bit
  misleading in this context. The procedure still applies without any
  change, and gives the right answer: $\cN=4$ super-Yang-Mills in
  $\cN=1$ notation, as it is easy to check from
  figures~\ref{fig:coniresolvdimer}b and \ref{fig:coniresolvdimer}c.}

\FIGURE{
  \includegraphics[width=0.9\textwidth]{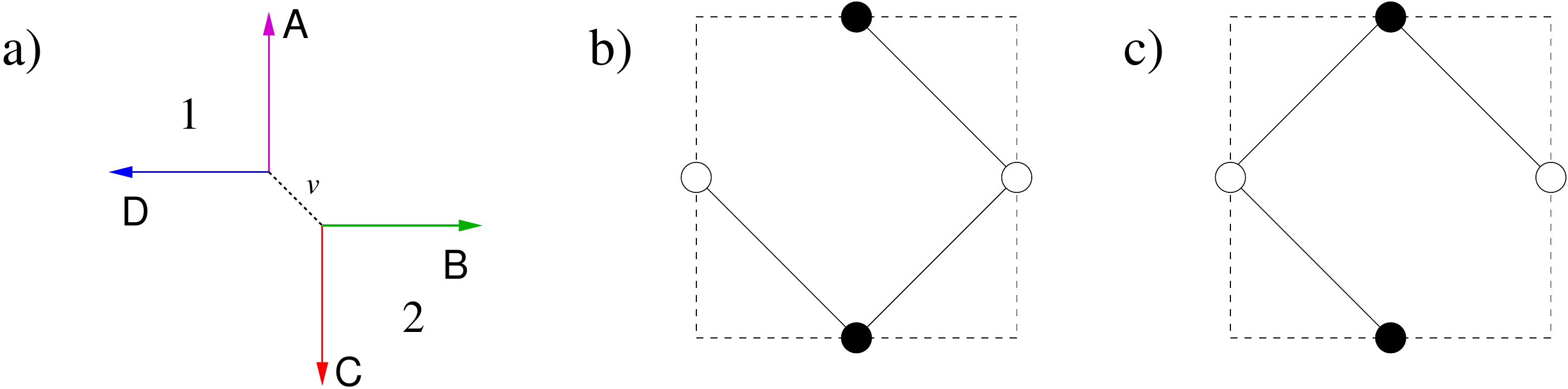}

  \caption{Resolution process for the conifold into two copies of flat
    space. a) Resolution in terms of the web diagram. The internal
    dashed lines denotes the blown up cycle, of size $v$. We have
    denoted the two daughter singularities 1 and 2, as in the text. b)
    Dimer model corresponding to the daughter singularity 1, obtained
    by removing the edge where B and C intersect (see
    figure~\ref{fig:conizz}a). c) Dimer model for the daughter
    singularity 2, obtained by removing the edge where A and D
    intersect. Note that both b) and c) match what one would obtain by
    directly applying the fast inverse algorithm to flat space.}

  \label{fig:coniresolvdimer}
}

The procedure we are describing can always be carried out for gauge
theories arising from D3-branes at a toric singularity.  In this case
every gauge group in the parent theory has rank $N$ while the gauge
factors at the two daughter singularities all have ranks $P$ and $Q$
with $P + Q = N$. If fractional branes are wrapped on any of the
collapsed cycles of the parent singularity, the resolutions of these
cycles do not lie on the moduli space and cannot be carried out.  In
other words, in the presence of fractional branes a partial resolution
can only be carried out if these branes are wrapped on cycles that are
entirely carried into one of the daughter singularities and not on the
cycle we are blowing up.  Fortunately, in this paper we will always go
in the other direction -- given daughter singularities with or without
fractional branes we will give a gluing prescription that will give a
larger parent singularity containing both of the daughters.  Thus, for
our purpose of describing the spectrum of the complete theory we can
assume that the blowup in question can be performed even in the
presence of fractional branes.

Given the gauge theory of the parent singularity,
and explicit expressions for the vevs of the bifundamentals responsible for the partial resolution,
we can give a complete description of the theory after the resolution,
including the massive mediator sector, with its couplings. There are
some simple rules for determining the massive spectrum, which apply even in the presence of fractional branes as long as the resolution is possible:

\begin{enumerate}
  % Set the labels in bold for this enumeration.
  \renewcommand{\labelenumi}{\bf \arabic{enumi}.}
\item Consider an edge which disappears in the $i^{th}$ daughter dimer
  diagram leaving a face with a gauge group of rank $K$.  Then there is also a massive 
  vector multiplet in the adjoint of
  $U(K)$ gauge factor.   This massive multiplet arises from Higgsing of a  $U(K) \times U(K)$ subgroup of the gauge factors of the two faces the
  edge used to separate in the parent theory.  Only the diagonal component of this subgroup survives the Higgsing, while the non-diagonal part gets a mass through the Higgs mechanism.
\item For each face in the parent dimer diagram, we obtain two
  massive vector multiplets in the bi-fundamental
  $(N_1,\overline{N_2})$ and its conjugate, where $N_{1,2}$ are the ranks of the faces in the daughter dimers that enclose the face in the parent.    (It might be that the faces in the daughters have expanded via the loss of edges, but they  never shrink in the procedure we are describing.)  
\item Consider an edge present in both daughter dimer diagrams separating faces with ranks $N_1^a, N_1^b$ and $N_2^a,N_2^b$ in the two daughters respectively.   Also suppose that the edge was oriented in the parent so that bifundamentals went from $a$ to $b$.   Then in the daughter theory there are massive 
  chiral multiplets in  $(N_1^a,\overline{N_2^b})$  and $(\overline{N_1^b},N_2^a)$ 
  bi-fundamental representations.   The dimer diagram ensures that globally,
  these types of chiral multiplets pair up consistently to form massive
  scalar multiplets \cite{GarciaEtxebarria:2006rw}. 
\item If the daughter dimer diagrams contain bi-valent nodes (nodes
  with two edges), the pair of edges gives a massive
  scalar multiplet in the bi-fundamental of the two faces 
  separate by the pair.
\end{enumerate}
If there are non-compact D7-branes threading the singularity, additional massive matter will appear (see section~\ref{sec:other-sectors} for details.)

\section{Joining singularities}
\label{sec:joining}

In the preceding section we discussed the process of partially
resolving a larger (mother) singularity into several smaller
(daughter) singularities. We now consider the reverse process, i.e.,
unresolving or joining daughter singularities to form a new
singularity. There are several reasons why this may be a more useful
way to proceed. First, from a bottom-up approach we start with a
low-energy theory that consists of several sectors, each of which is
described by putting D-branes on a singularity. We then want to know
how these singularities fit together, such that we can determine the
massive spectrum of messenger particles.  In addition, the process of
splitting singularities described in
section~\ref{sec:partial-resolution} has an ambiguity in the case that
the web diagram of the parent geometry has parallel external legs of
the same orientation that get split. In such a case one is free to
choose which zig-zag path goes to which side of the split, and the
resolution procedure we described in
section~\ref{sec:partial-resolution} will give different answers
depending on our choice. This phenomenon is a manifestation of Seiberg
duality \cite{Seiberg:1994pq}, which in this context appears as toric
duality \cite{Feng:2000mi}. More generally, there exist different
Seiberg-dual dimer models describing the same geometry. (Note that the
conifold is special since taking the Seiberg dual of any face of the
dimer brings us back to the same dimer, i.e., it has a single Seiberg
phase).

This effect is important when building particular models, since upon
resolving the daughter singularity in different ways (thus getting
Seiberg dual daughters), or upon starting with a different Seiberg
dual phase of the mother theory, one obtains different daughter
theories.  The latter are usually different from the theories one is
interested in putting together.  In this section we describe a
procedure, a simple combination of the fast inverse algorithm
described in section~\ref{sec:fast-inverse} and the partial resolution
algorithm described in section~\ref{sec:partial-resolution}, that
allows us to join any two singularities in such a way that upon
resolution we obtain the two desired dimer models for the two daughter
singularities.

\subsection{The process  of unresolving}

Let us start with the dimer models that we want to join. For
illustration, we reconstruct the conifold theory out of two copies of
flat space. The complete procedure is shown in
figure~\ref{fig:conijoin}. This particular example does not exhibit
any of the subtleties having to do with Seiberg duality, but the
examples in section~\ref{sec:dark-mssm} do, and they can be analyzed
in the same way. In the general case, in addition to specifying the
toric singularities that we want to join, we also should choose the
particular dimer model representative of the daughter singularity that
we want to join.

\FIGURE{
  \includegraphics[width=0.9\textwidth]{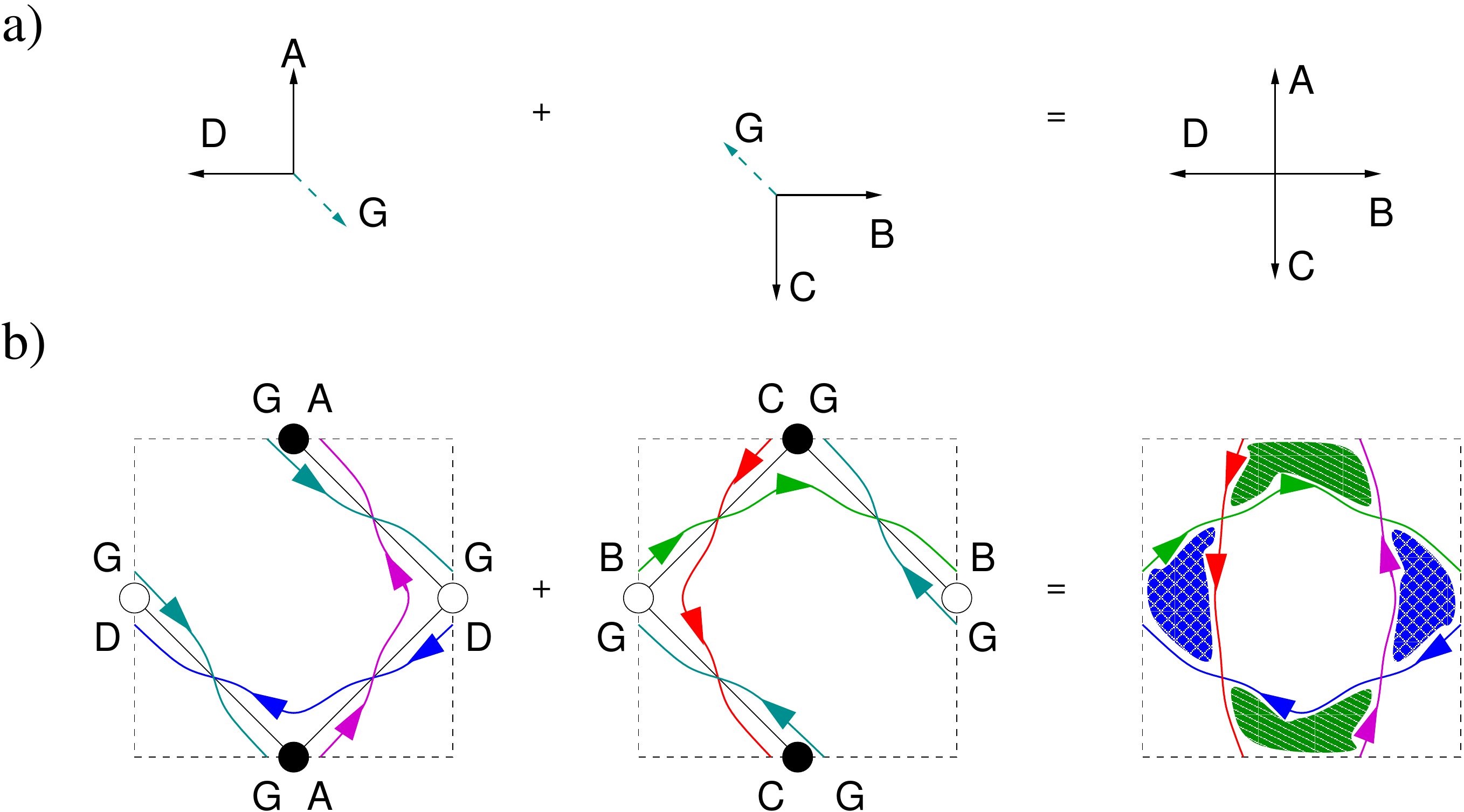}

  \caption{Joining two copies of flat space along one external leg. a)
    Web diagram perspective. We have indicated the external edge to be
    joined (G) by the dashed colored line. b) Dimer model
    perspective.}

  \label{fig:conijoin}
}

Once we have picked the singularities, we need to choose along which
external leg of their web diagram we want to join them. In general, we
might need to perform a $SL(2,\bZ)$ transform of the daughter dimer
models in order to align the legs that we want to join.  This is
always possible if the legs are primitive vectors (i.e., for each
vector the components are relatively prime).  If a leg is
non-primitive, we write it as a multiple of a primitive vector and
then the web diagram can only be joined to another diagram with a leg
that is the same multiple of an oppositely oriented primitive vector.
In this case, in addition to the two sectors we want to join, we might
have to attach additional daughter singularities in order to ensure
that the combined parent singularity remains toric (we will see
examples of this later). The new sector will not contribute massless
degrees of freedom if we do not put any brane at the corresponding
daughter singularity, but it will modify the precise form of the
massive mediators and their couplings to massless modes.

The next step in our graphical algorithm for joining theories is to
remove from each daughter dimer model everything except the external
zig-zag paths that survive in the parent.\footnote{The resulting
  diagram is closely related to what is referred to as a
  \emph{harlequin diagram} in \cite{Franco:2007ii}, the only
  distinction being that we remove from the harlequin diagram the
  zig-zag that is going to join the singularities.} In particular,
this removes the zig-zag paths corresponding to the daughter legs that
we are going to join. In the case of the conifold in
figure~\ref{fig:conijoin}, this means leaving only the A,D zig-zag
paths in the first dimer model, and B,C in the second dimer model,
removing all nodes, edges, and and the zig-zag path G.

Finally, we superpose the two sets of remaining zig-zag paths, and
compute the new theory as in the inverse algorithm. By construction,
the new theory descends to the two dimer models we started with upon
partial resolution.\footnote{There is still some ambiguity in how to
  superpose the two sets of zig-zag paths. Namely, we could deform the
  zig-zag paths in such a way that we induce no Yang-Baxter transform
  (i.e. Seiberg duality coming from continuation across triple
  intersections of zig-zag paths \cite{Hanany:2005ss}) in any of the
  daughter dimer models, but we induce one such Yang-Baxter transform
  in the mother theory. This is perfectly consistent, and reflects the
  fact that even if the daughter singularities have their Seiberg dual
  phase fixed, this does not fix uniquely the Seiberg dual phase of
  the mother theory.}

In this graphical algorithm we started by removing all the edges and nodes in the daughter dimer, and then reconstructed the dimer from the zig-zag paths of the parent theory produced by the joining algorithm.  Of course we could have simply left the daughter edges and nodes in, since the joining algorithm essentially introduces a few new bifundamentals (edges) and superpotential terms (nodes).  It is simply easier as a graphical technique  to proceed as we described, and to reconstruct the dimer
model for the joined theory from scratch.

\subsection{Tadpole cancellation}
\label{sec:tadpoles}

The previous algorithm works without any further subtleties in the
case that all faces in the final (joined) dimer model have the same
rank. Note that this condition is equivalent to each subsector having
all its faces of the same rank (although different subsectors can have
different ranks). When all faces have the same rank our dimer model
defines a superconformal field theory \cite{Franco:2005rj}. For model
building applications we  require theories that are
non-conformal. We can achieve this in two ways: (1) by putting fractional
branes in the singularity, and/or (2) by introducing non-compact D7 branes in
the geometry. Let us analyze each of these possibilities in turn.

The first possibility consists of choosing the ranks of the faces in
the dimer model to be different. This situation is usually referred to
as putting \emph{fractional branes} in the system, since this kind of
rank assignment comes from wrapping D5 and D7 branes on the vanishing
cycles (the configuration with all ranks equal corresponds to putting
a stack of D3 branes at the singularity).

In such a configuration local tadpole cancellation becomes an
important issue. The condition for tadpole
cancellation in $\cN=1$ quiver gauge theories is that every node in
the quiver has the same number of fundamental and antifundamental
chiral multiplets.\footnote{This condition is the same one as anomaly
  cancellation of the resulting gauge theory, with the addition that
  empty nodes must be anomaly free too. A simple way of understanding
  this extra condition is that we can ``fill in'' the empty nodes
  without changing the compact tadpoles by bringing a D3 brane into the
  singularity. This adds one unit to the rank of all faces,
  reducing the check for consistency to usual anomaly cancellation.}
It is easy to see that this holds in dimer models with faces of equal
rank: recall that in a dimer model, due to the alternating black/white
coloring of the nodes, each face has to have an even number of edges,
with alternating orientations, so each face has as many fields in the
fundamental as in the antifundamental.

Once the ranks of the gauge groups are not equal, tadpole cancellation
is no longer guaranteed, and we need to check it for each rank
assignment. However, as long as the theories that we join are tadpole
free, the joined singularity appears to be tadpole free too.
Heuristically, the reason why it is enough to satisfy tadpoles in each
subsector is that the compact cohomology charge adds simply when we
join singularities. However, we do not have a complete proof of this
observation, and it is certainly no longer true if we add non-compact D7
branes into the problem, which is often necessary to construct
realistic models.  In this case, anomaly cancellation in the joined
theory must be checked by hand.

The reason for this is that the requisite D7 branes naturally extend along the
external legs of the web diagram. When we join singularities we might
be forced, to have consistency of the construction, to introduce D7 branes
along subsectors that did not have them originally. As an example, let
us consider sectors $A$ and $B$ where $A$ had some D7 branes, while $B$ did
not. Upon joining $A$ with $B$, the D7 branes of $A$ might extend
along the leg that connects $A$ to $B$ and thus  have to exit the web diagram along an external leg of $B$.
This introduces new flavors charged under the fractional branes
in $B$. In general this also introduces local tadpoles in the theory
$B$. More geometrically, the D7 brane has local tadpole charge under
the compact homology of the new subsector, and we need to cancel this by
adding extra fractional D-branes in the $B$ sector.

Thus, we are forced to change sector $B$ by changing
the ranks of the gauge groups, and by adding extra flavors to it. In
some cases this modified model might be just as good
phenomenologically as the original.  In general
the D7 tadpole condition introduces an important constraint that the
subsectors must satisfy.  Even if the constraint is
important, it is certainly possible to build interesting models that
satisfy it (see \cite{GarciaEtxebarria:2006rw} for some examples).

Finally, having obtained the tadpole free dimer model for the parent
theory resulting from joining daughter singularities, we are left to
determine the ranks of the gauge groups in the parent related to the
ranks in the daughters.  Every face in the parent dimer diagram
descends to part of face in each daughter diagram (the faces in the
daughter arise from removing some edges in the parent).  The rank
associated to the parent face is the sum of the ranks of daughter
faces that it participates in.

\section{A simple dark matter model}
\label{sec:dark-mssm}

With these tools at hand, we can construct local models
of dark matter.  Our constructions are motivated by simplicity, to illustrate our methods, and
phenomenologically better choices certainly exist.  Because our constructions are modular, such improvements are easily incorporated, as long as they are
toric. We present a number of different alternatives to this model in
section~\ref{sec:other-sectors}.

Our model consists of three sectors, coupled by massive mediators: a
supersymmetry breaking sector, a ``MSSM'' visible sector, and a dark
matter sector, which we take to be a copy of the MSSM sector, with the
only difference that it couples differently to the supersymmetry
breaking sector. We show the structure of the resulting theory in
figure~\ref{fig:moose}.

\FIGURE{
  \includegraphics[width=0.8\textwidth]{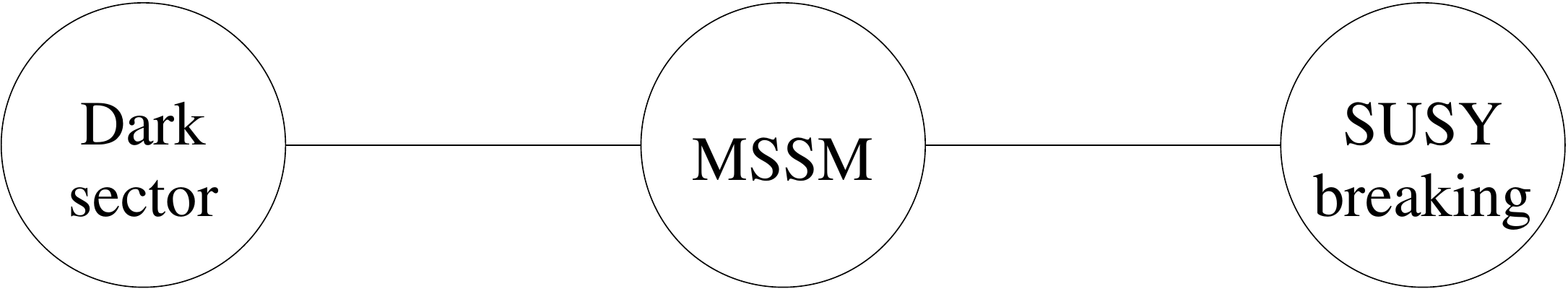}

  \caption{Schematic representation of our simple model. The nodes
    represent the different subsectors of light degrees of
    freedom. The lines connecting the different nodes represent
    massive multiplets charged under the different nodes.}

  \label{fig:moose}
}

We achieve this using the techniques described in previous sections:
we locally engineer each of the sectors as a dimer model, and use
toric techniques to paste them together into a slightly resolved
bigger singularity. In fact, a big part of the work has already been
done for us in \cite{GarciaEtxebarria:2006rw}, where a local
configuration with two sectors implementing gauge mediated
supersymmetry breaking was described. We explain how to attach a
third sector containing dark matter to the theory found in
\cite{GarciaEtxebarria:2006rw}.

In more detail, our visible sector is of the trinified form ($SU(3)^3$
gauge group), and the dark matter sector is an exact copy of the
standard model, as in the well studied mirror world proposal of
\cite{Foot:1995pa} (adapted for trinification). We take the
supersymmetry breaking sector to be a theory with a runaway potential
and a metastable vacuum at the origin of moduli space. In
section~\ref{sec:improved-breaking} we will replace this supersymmetry
breaking sector by a better behaved geometry.

The resulting model is of the ``minimal superdark moose'' form
described in \cite{ArkaniHamed:2008qp}.  In such a model, in order to
get weak scale dark matter we want the same mechanism to set the
$\mu$-term in the MSSM and the dark sector.  Thus, the $\mu$-term has
to be generated by a common mechanism (such as a D-brane instanton
\cite{Blumenhagen:2006xt,Ibanez:2006da,Buican:2006sn,Ibanez:2007tu})
that can appear in the local geometries of both the dark sector and
MSSM, rather than by RG running of the interactions with the SUSY
breaking sector. Our toy models will not generate the $\mu$ term
non-perturbatively, but in other local models studied in the
literature this can be arranged. See for instance \cite{Ibanez:2007tu}
for a explicit set of semi-realistic models from branes at toric
singularities in which D-brane instantons generate the $\mu$ term
non-perturbatively.

\subsection{Supersymmetry breaking sector}
\label{sec:susy-breaking}

One simple way to obtain supersymmetry breaking in local models is to
introduce an appropriate kind of fractional brane in the local
geometry. The resulting theory is no longer conformal, and in
particular infrared dynamics generate a runaway superpotential that
breaks supersymmetry. These kinds of branes (called \emph{DSB} branes
in the classification of \cite{Franco:2005zu}) are very generic, the
simplest example appearing already in the complex cone over
$dP_1$. For the moment we take this geometry, with the appropriate
brane, to be our supersymmetry breaking sector (we will improve the
problem of having a runaway potential in following sections). The
relevant data for this sector is shown in
figure~\ref{fig:fractionaldp1}.

\FIGURE{
  \includegraphics[width=0.9\textwidth]{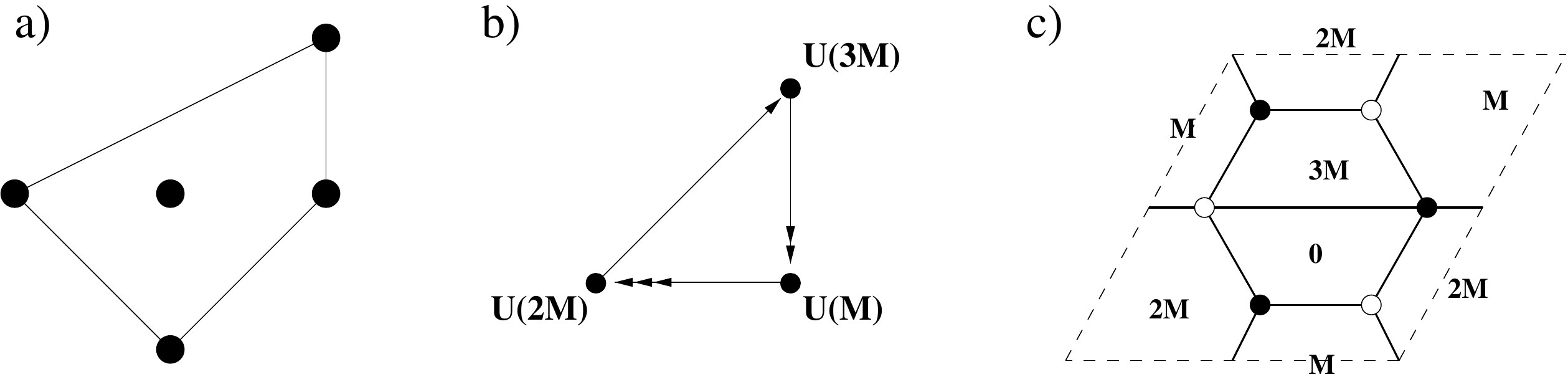}

  \caption{a) Toric diagram for the complex cone over $dP_1$. b) The
    quiver gauge theory for $M$ DSB branes in the complex cone over
    $dP_1$. c) Dimer model.}

  \label{fig:fractionaldp1}
}

Let us review why this theory has a runaway potential
\cite{Franco:2005zu,Intriligator:2005aw}. We  denote the fields
going from $U(M)$ to $U(2M)$ as $X_{12},Y_{12},Z_{12}$, the fields
going from $U(3M)$ to $U(M)$ $X_{31},Y_{31}$ and the field going from
$U(2M)$ to $U(3M)$ $X_{23}$. We can read off the superpotential for the theory
from the dimer model in
figure~\ref{fig:fractionaldp1}c:
\begin{equation}
  W = X_{23}X_{31}Y_{12} - X_{23}Y_{31}X_{12}.
\end{equation}
Note in particular that $Z_{12}$ is decoupled, and parameterizes a
flat direction. Since this is only a toy model, we will ignore this
issue in the following.

As we remarked in footnote~\ref{U(1)-footnote}, generically the $U(1)$
factors  get masses by mixing with background RR fields. In
the particular case we are dealing with  all $U(1)$
factors are anomalous (except for the overall $U(1)$, which
decouples), and thus necessarily get a mass. So we are in fact dealing
with a $SU(3M)\times SU(2M)\times SU(M)$ gauge theory.

Let us assume that the node of rank $3M$ confines first. In this case
the theory develops an ADS superpotential, given by
\cite{Berenstein:2005xa,Franco:2005zu,Bertolini:2005di,Intriligator:2005aw}
\begin{equation}
  W = M_{21} Y_{12} - M'_{21} X_{12} +
  M\left(\frac{\Lambda_3^{7M}}{\det \cM}\right)^{\frac{1}{M}},
\end{equation}
where we have introduced the mesons $M_{21}=X_{23}X_{31}$,
$M'_{21}=X_{23}Y_{31}$, and the mesonic $2M\times 2M$ matrix
$\cM=(M_{21};M'_{21})$.
We can easily see the runaway direction from here: the F-term
equations for $Y_{12}$ and $X_{12}$ want to set
$M_{21}=M'_{21}=0$. But now the F-term equation for the mesons pushes
$Y_{12}$ and $X_{12}$ to infinity, due to the inverse power of the
meson matrix appearing in the ADS term.

It is possible to add a metastable vacuum at the origin of the runaway
direction in moduli space by adding vector-like massive matter to our
theory \cite{Franco:2006es}, see also \cite{GarciaEtxebarria:2007vh}
for generalizations to a large class of toric singularities. A simple
way of achieving this is by introducing non-compact D7 branes that
come close to the singularity. Open strings between the D3 branes at
the singularity and the flavor D7 branes give rise to massive
bifundamentals.

In our setup, we have another natural candidate for the massive
flavors -- the open strings between the supersymmetry breaking sector
and the visible sector. This configuration is an interesting variation
of the direct mediation \cite{Affleck:1984xz,Luty:1998vr} family of
models. (See also \cite{Intriligator:2006dd,Kitano:2006xg} for early
studies of direct mediation of metastable supersymmetry breaking.)

\subsection{The visible/dark matter sector}

For the visible and dark matter sectors we choose to put our branes at
the singular point of the $\bC^3/\bZ_3$ orbifold, also known as the
complex cone over $dP_0$. This gives the theory described in
figure~\ref{fig:dp0}. These kinds of models with gauge group $SU(3)^3$
and three families of bifundamentals are usually referred to as
\emph{trinification} models (see for example
\cite{Willenbrock:2003ca}). The idea of taking the dark matter sector
to be a ``mirror world'', namely a copy of the standard model sector,
also has a long history in phenomenology and it is a well studied
possibility \cite{Foot:1995pa}.

\FIGURE{
  \includegraphics[width=0.9\textwidth]{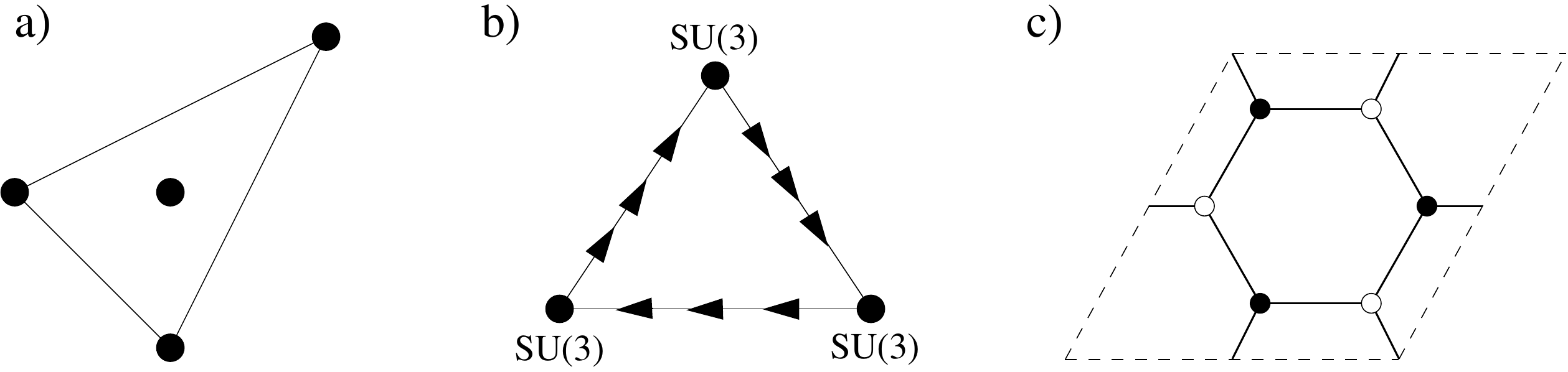}

  \caption{a) Toric diagram for the complex cone over $dP_0$. b) The
    quiver gauge theory for 3 D3 branes at the singular point of
    $\bC^3/\bZ_3$. c) Dimer model, all faces have rank 3.}

  \label{fig:dp0}
}

\subsection{Putting the different sectors together}

As we mentioned in the introduction to this section, part of the work
of combining the three sectors was already done in
\cite{GarciaEtxebarria:2006rw}, where one $dP_1$ and one $dP_0$ were
joined together into a $X^{3,1}$ singularity, studied in
\cite{Hanany:2005hq,Franco:2005rj}, see figure~\ref{fig:X31}.  We are
left to consider the last step of joining the $X^{3,1}$ singularity
with the complex cone over $dP_0$. From the latter, we then get a new
dark matter sector, given by a copy of the MSSM.

\FIGURE{
  \includegraphics[width=0.9\textwidth]{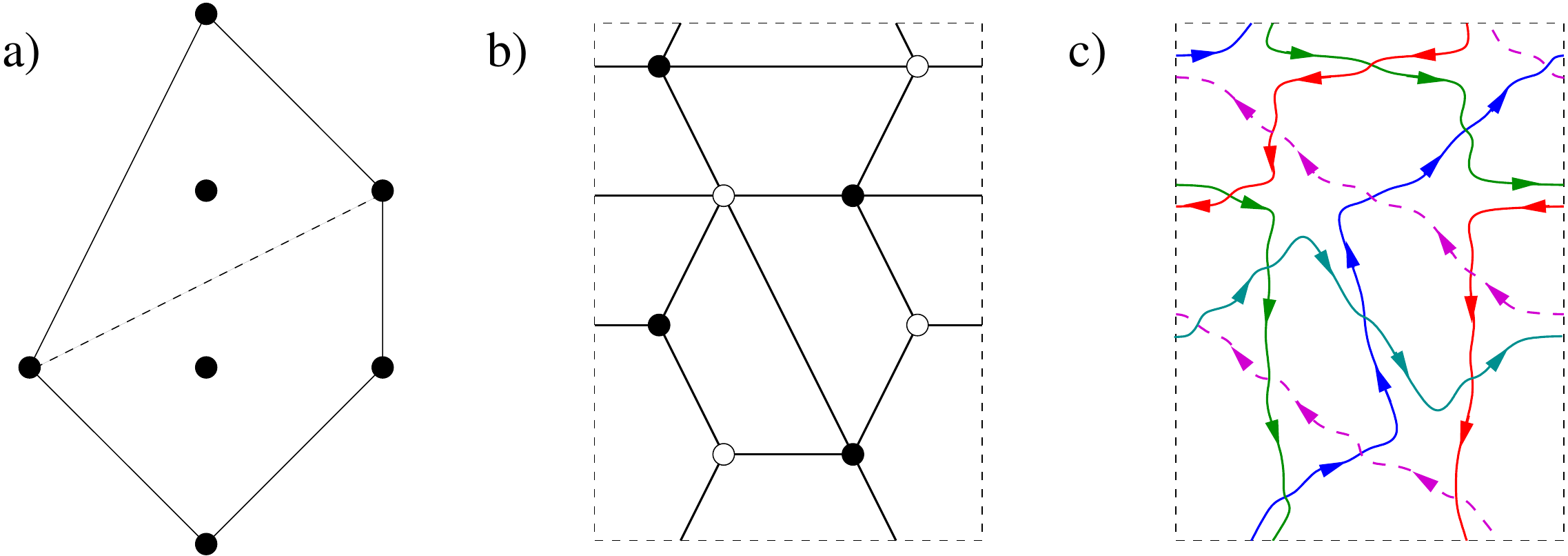}

  \caption{a) Toric geometry for the $X^{3,1}$ singularity, obtained
    by joining the $dP_0$ and $dP_1$ singularities along the dashed
    line. b) Dimer model. c) Harlequin diagram obtained from the dimer
    model. We have indicated with a dashed magenta line the zig-zag
    path along which we  join $X^{3,1}$ to the $\bC^3/\bZ_3$
    geometry.}

  \label{fig:X31}
}

Following the discussion in section~\ref{sec:joining}, this is done by
joining technique, placing the complex cone of $dP_0$ at a finite
distance from the $X^{3,1}$ singularity. Let us consider the $X^{3,1}$
subsector. The corresponding dimer model is given in
figure~\ref{fig:X31}b. We draw the zig-zag paths of this model in
figure~\ref{fig:X31}c. In order to join the singularities, we need to
remove the path corresponding to the internal leg (shown with a dashed
line in figure~\ref{fig:X31}c), and superpose the external legs coming
from the $\bC^3/\bZ_3$ part of the geometry that we are attaching. As
shown in figure~\ref{fig:dm-simple}a, the external legs have slopes
$(-1,-1)$ and $(-1,2)$. In figure~\ref{fig:dm-simple}b we have
superposed a couple of zig-zag paths with those winding numbers to the
amputated harlequin diagram obtained from
figure~\ref{fig:X31}c. Finally, in figure~\ref{fig:dm-simple}c we have
reconstructed the final dimer model from the set of zig-zag paths, as
described in section~\ref{sec:fast-inverse}.

\FIGURE{
  \includegraphics[width=0.9\textwidth]{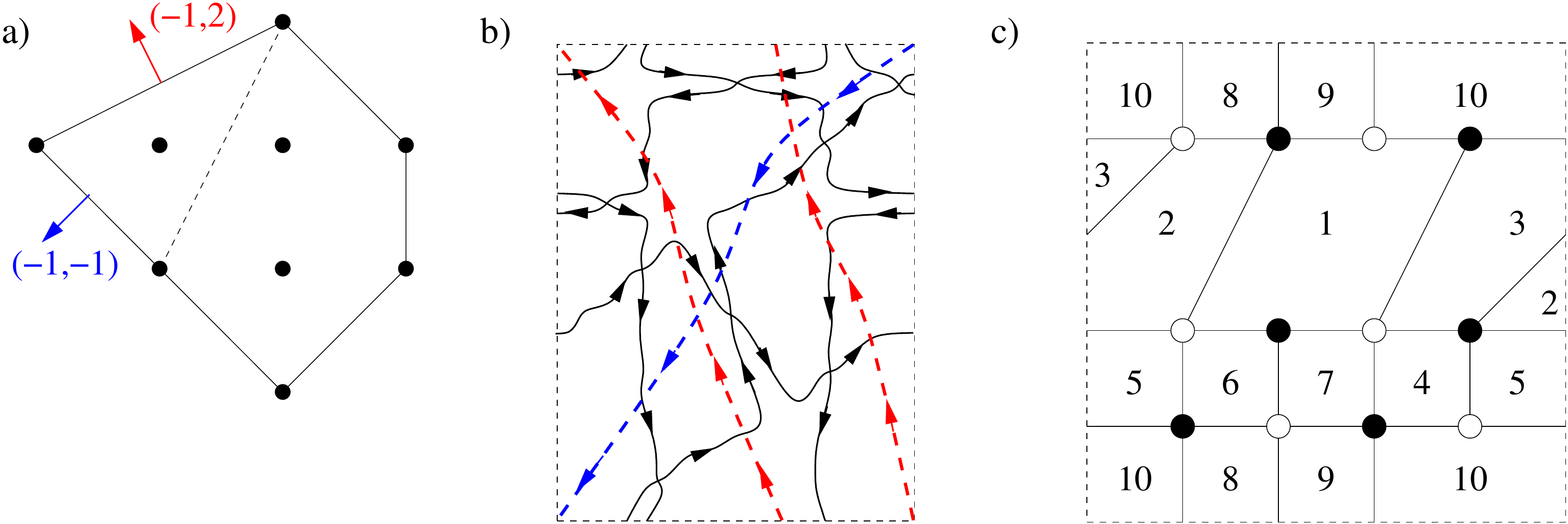}

  \caption{a) Toric geometry for the three sector model, obtained by
    joining the $X^{3,1}$ geometry to $dP_0$ along the dashed line. We
    have shown the external legs of the $dP_0$ sector that we are
    attaching. b) Harlequin diagram for the joined theory, obtained by
    adding the external zig-zag paths of $\bC^3/\bZ_3$ (dashed, in
    color) to the external zig-zag paths in harlequin diagram of
    figure~\ref{fig:X31}c (in black). c) Reconstructed dimer model. We
    have numbered the faces for later reference.}

  \label{fig:dm-simple}
}

We have included fractional branes in order to obtain a supersymmetry
breaking sector, so in order to completely specify the model we also
need to specify the ranks of the different faces in the dimer
model. This is simple to determine once we have the explicit
description of the resolution in terms of dimer models (see
section~\ref{sec:tadpoles}).  Each face of the mother dimer model maps
to a specific face of the daughter dimer models. The rank of the face
in the mother theory is simply the sum of the ranks of the
corresponding faces in the daughter theory. That is, we have the
following expression for the rank of the face $i$ in the mother dimer
model:
\begin{equation}
  rank(F_i) = \sum_{daughter\, d} rank(F_d(F_i))
\end{equation}
where $F_d(F_i)$ denotes the face in the daughter dimer model
corresponding to $F_i$ in the mother.

The procedure is systematic, so let us just quote the result that we
get. Numbering of the faces as in figure~\ref{fig:dm-simple}c, the
rank vector is given by:
\begin{equation}
  \vec{Q} = 6\cdot(1,1,1,1,1,1,1,1,1,1) + (M,2M,M,M,0,3M,3M,2M,3M,2M).
\end{equation}
That is, our theory has gauge group $SU(6+M)\times SU(6+2M)\times
SU(6+M)\times \ldots$. The first term comes from the regular branes
at the $dP_0$ singularities, corresponding to the standard model and
the dark matter sector (they contribute 3 D3 branes each), and the
second term comes from the fractional brane at the $dP_1$
singularity. It is a simple exercise to check that there are no
tadpoles with this assignment of ranks.

\section{Some interesting modifications of the construction}
\label{sec:other-sectors}

In the previous section we have discussed the construction of a
particular dark matter model. Our construction is highly modular, and
this section will illustrate this point by discussing some interesting
modifications of the structure given above. We have chosen examples
that require introducing some extra ingredients into our construction.

\subsection{Flavor D7 branes and an improved standard model}

\FIGURE{
  
\ifpdf
  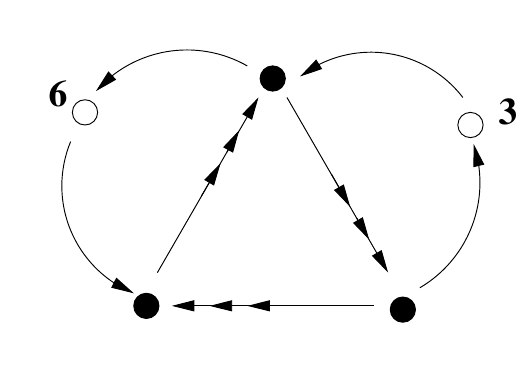
  \else
  \input{dp0mssm.pstex_t}
\fi

  \caption{A better toy model for the MSSM, from
    \cite{Aldazabal:2000sa}. The filled dark dots denote gauge groups,
    while the white dots denote global symmetry groups, coming from
    non-compact D7 branes.}

  \label{fig:dp0mssm}
}

We can improve the visible sector using the model in
\cite{Aldazabal:2000sa},\footnote{A number of semi-realistic models
  involving branes at toric singularities and non-compact D7 branes
  have been recently described in \cite{Conlon:2008wa}.} which is
given by a fractional brane assignment in the $\bC^3/\bZ_3$ orbifold,
with some non-compact D7 branes with baryonic charge. The quiver for
this theory is shown in figure~\ref{fig:dp0mssm}. It includes the
usual three family MSSM (without right handed neutrinos), and some
extra vector-like matter. Here hypercharge does not correspond to the
$U(1)$ node (which is anomalous), but rather to a non-anomalous
combination of the abelian factors of the gauge symmetry on the
nodes. In particular, it is given by:
\begin{equation}
  Y = \frac{1}{3}Q_3 + \frac{1}{2}Q_2 + Q_1
\end{equation}
where $Q_N$ represents the abelian part of the $U(N)$ node.

\medskip

Let us review for completeness how to introduce flavor in our theories
using non-compact D7 branes passing through the singularity. In the
following we will discuss massless flavors, but it is also possible to
discuss massive flavors by introducing D7 branes that do not touch the
singularity \cite{Karch:2002sh}. This is very convenient for building
models of metastable supersymmetry breaking, see for example
\cite{Franco:2006es,GarciaEtxebarria:2007vh}.

\subsubsection*{D7 branes and dimer models}

Let us start by briefly reviewing how to represent non-compact D7
branes in the language of dimer models \cite{Franco:2006es}.

\FIGURE{
  
\ifpdf
  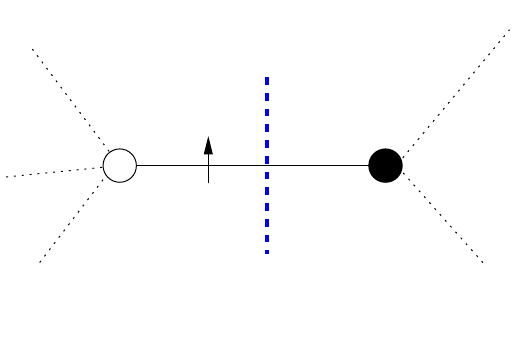
  \else
  \input{D7-dimer.pstex_t}
\fi

  \caption{Local representation of a D7 brane, shown as a dashed blue
    line. $SU(N_1)$ and $SU(N_2)$ denote a couple of faces in the
    dimer model, $X_{21}$ is an edge connecting them, and the black
    and white nodes denote superpotential terms. The dotted lines
    represent extra edges in the dimer that do not enter our
    discussion.}

  \label{fig:D7-dimer}
}

Non-compact D7 branes wrap a non-compact divisor of the toric
geometry. In terms of the web diagram, the basic class of non-compact
divisors are associated with external legs. The basic idea is that we
can think of the basic divisor $z_i=0$, where $z_i$ is a field in the
gauge linear sigma model (GLSM), as determined by the couple of
external legs of the web diagram bounding it.\footnote{In general,
  apart from the geometric divisor class of a D7 brane, we also need to
  specify the value of possible Wilson lines on it. The dimer model
  also captures this information in a natural way.}

In the mirror of the toric variety, this D7 brane naturally projects
to a curve wrapping a one cycle in a certain Riemann surface. We will
explore this Riemann surface extensively in section~\ref{sec:mirror},
but let us postpone that discussion for now. The basic point is that
this Riemann surface naturally encodes the dimer \cite{Feng:2005gw},
and from the mirror description of the non-compact D7 brane we can read its
description in dimer model terms. The end result is simple to state:
D7 branes wrapped on basic divisors can be introduced as decoration in
the dimer model given by a line going from a face to a neighboring
face, crossing one bifundamental edge.\footnote{There is a simple
  generalization for D7 branes wrapping general divisors. They get
  mapped to general open paths in the dimer model
  \cite{Imamura:2008fd,Forcella:2008au}.}  We have shown this in
figure~\ref{fig:D7-dimer}.

The D7 brane  introduces a couple of chiral multiplets
$X_{17},X_{72}$ into the theory, charged in the bifundamental of the
flavor D7-brane group and $SU(N_{(1,2)})$. There is also a new
superpotential term given by:
\begin{equation}
  W_{D7} = X_{72} X_{21} X_{17}
\end{equation}
where we are omitting the overall trace, as usual.

The improved MSSM given above is easily described in this language. We
refer the reader to \cite{GarciaEtxebarria:2006rw} for details of the
final dimer model, and a two-sector GMSB model involving one such
subsector.

\subsection*{D7 branes and partial resolution}
Upon partial resolution of the singularity the D7-branes contribute
additional matter.  This gives one more massive messenger rule in
addition to the four rules given in section~\ref{sec:partial-resolution}:

\begin{enumerate}
\item[\bf{5.}] For each D7-brane passing through an edge of type 1
  there is a massive scalar multiplet in the fundamental
  representation of the $U(N_2)$ gauge factor corresponding to the
  resulting recombined face. Similarly for edges of type 2. When there
  are $N_7$ D7 branes across such an edge the massive multiplet
  transforms as $(N_{D3}, \overline{N_7})$.
\end{enumerate}

\subsection{Orientifolds and improved supersymmetry breaking}
\label{sec:improved-breaking}

The supersymmetry breaking sector described in the previous section
has a couple of important shortcomings. First of all, it has a
decoupled flat direction $Z$, which generically  gets a mass after
supersymmetry breaking (starting with two loops, since it is decoupled
at the level of the superpotential), and might conceivably destabilize
the meta\-stable vacuum. Another important issue is that there is a
runaway direction in the potential. While the metastable vacuum can be made
long-lived, it is preferable to have a theory with a bona fide
stable vacuum, in addition to the metastable one. In fact, various
such configurations exist in the context of branes at singularities,
see \cite{Argurio:2006ny,Argurio:2007qk,Buican:2007is,Amariti:2008es}
for some simple examples that fit naturally in our toric framework. In
this section we will show how we can replace the supersymmetry
breaking sector in section~\ref{sec:dark-mssm} with the one proposed
in \cite{Argurio:2007qk}. Other cases can be shown to work similarly.

\FIGURE{
  \includegraphics[width=0.8\textwidth]{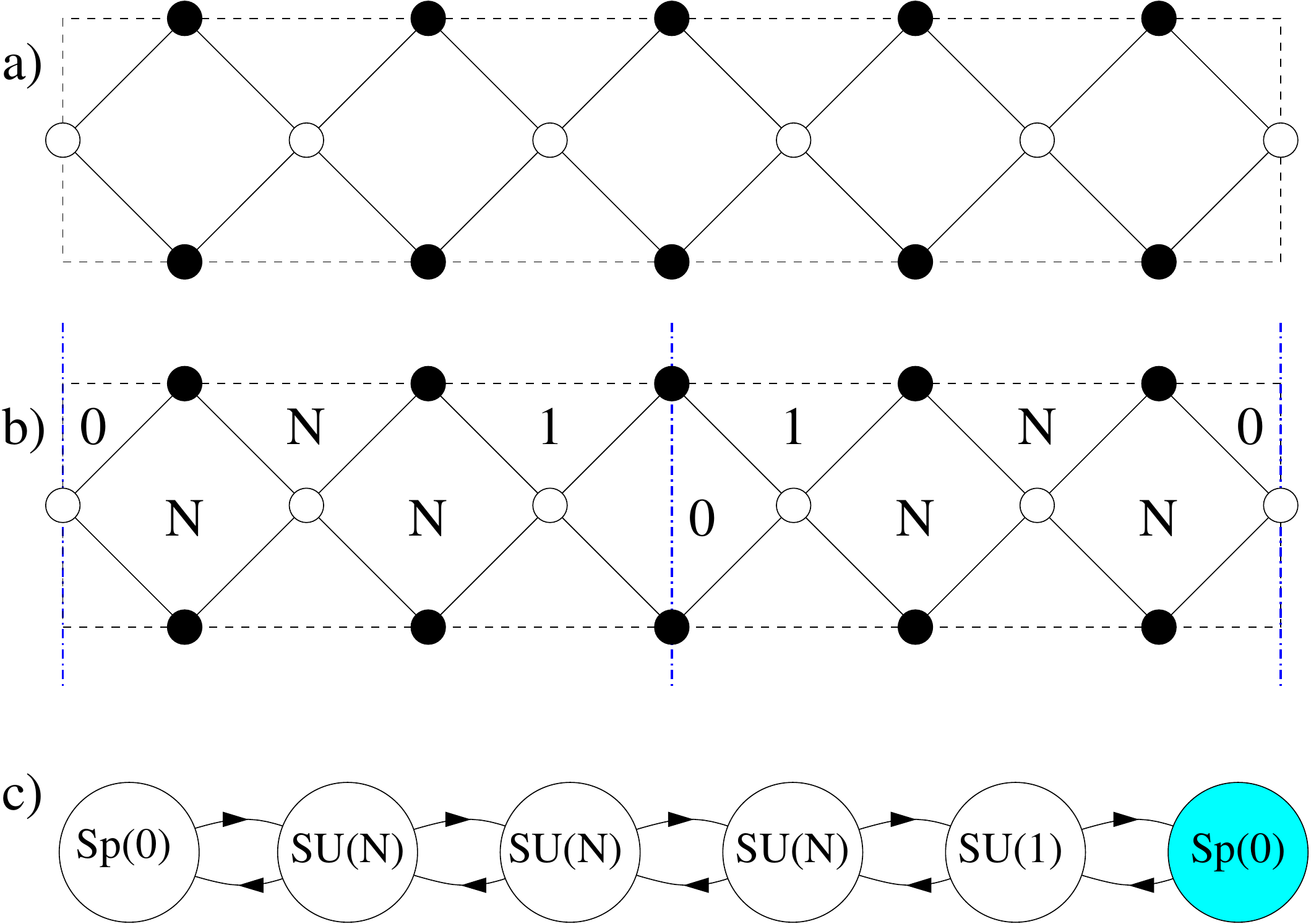}

  \caption{a) Dimer model for the $\bZ_5$ orbifold of the conifold. b)
    Line orientifold \cite{Franco:2007ii} of the orbifold, giving the
    model used in \cite{Argurio:2007qk}. We have denoted the
    orientifold plane by the vertical blue dashed line. We have also
    indicated the ranks of the faces. c) Quiver representation of the
    theory. The rightmost (colored) $Sp(0)$ node  gives rise to a
    mass term through a stringy instanton effect.}

  \label{fig:orbiconi}
}

Let us start by recalling the proposal of
\cite{Argurio:2007qk}. Consider a $\bZ_n$ orbifold of the conifold,
with $n \geq 5$. We  take $n=5$, for simplicity. The relevant
dimer model is shown in figure~\ref{fig:orbiconi}a. We  also
introduce an orientifold plane in the configuration, as shown in
figure~\ref{fig:orbiconi}b. In \cite{Argurio:2007qk} it was shown that such a configuration gives
rise to a metastable long-lived ISS vacuum~\cite{Intriligator:2006dd}, in addition to the usual
supersymmetric vacuum.

We  choose the charges of the orientifold planes and the ranks of
the faces such that we have an $Sp(0)\times SU(N)\times SU(N)\times
SU(N) \times SU(1) \times Sp(0)$ gauge theory with vector-like
matter. The different factors in the gauge group require some
explanation. The $SU(1)$ gauge group denotes a face where there is a
single fractional D3 brane. The $Sp(0)$ factor comes from a face
mapped to itself under the orientifold action. The sign of the
orientifold is chosen in such a way that it gives rise to $Sp(M)$
factors in the gauge theory. In this example we  put no fractional
branes on the face, so we end up formally with a $Sp(0)$ factor. As it
is well understood by now from a variety of different viewpoints (see
for example
\cite{Intriligator:2003xs,Argurio:2007qk,Ibanez:2007rs,Argurio:2007vqa,Bianchi:2007wy,Aganagic:2007py,GarciaEtxebarria:2008iw},
or \cite{Blumenhagen:2009qh} for a review), $Sp(0)$ gauge groups give
rise to non-perturbative dynamics in string theory due to $O(1)$
D-brane instantons wrapping the $Sp(0)$ node. In our case, the effect
of the instanton is to give a small mass to the bifundamental fields
between the $SU(1)$ node and its neighboring $SU(N)$ node.

Let us see how we can construct a three sector model with this
geometry as the supersymmetry breaking sector, replacing the complex
cone over $dP_1$. Our first task is to understand how the orientifold
acts on the geometry. We can determine this as follows. Consider the
following GLSM description of the $\bZ_5$ orbifold of the conifold:
\begin{equation}
\begin{tabular}{l|cccccc}
  & $a$ & $b$ & $c$ & $d$ & $e$ & $f$\\
  \hline
  $\bC^*$ & 1 & 1 & -1 & -1 & 0 & 0\\
  $\bC^*$ & 0 & 1 & 0 & 4 & -5 & 0\\
  $\bC^*$ & 4 & 0 & 1 & 0 & 0 & -5
  \label{table:Z5-GLSM}
\end{tabular}
\end{equation}
We show the resulting toric diagram in
figure~\ref{fig:Z5coni}. Note that the external nodes of the
toric diagram, which correspond to non-compact divisors of the toric
geometry,  are associated with fields in the GLSM. As an example, the upper right hand node
of the toric diagram in figure~\ref{fig:Z5coni} corresponds to the
$b=0$ divisor.

\FIGURE{
  
\ifpdf
  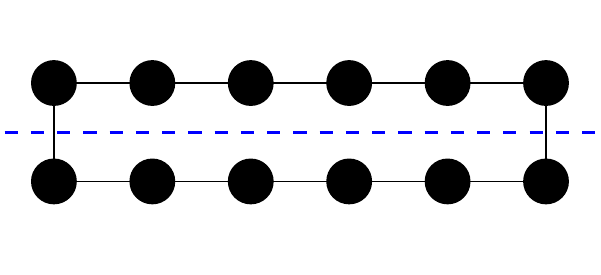
  \else
  \input{Z5coni.pstex_t}
\fi

  \caption{Toric diagram for the $\bZ_5$ orbifold of the conifold. The
    dashed blue line denotes the action of the orientifold action on
    the geometry. We have also indicated which non-compact divisors of
    the toric geometry are associated with the fields in the GLSM.}

  \label{fig:Z5coni}
}

The gauge invariants describing the geometry are:
\begin{equation}
  \begin{split}
    x & = a^5 d^5 e^4 f^4\\
    y & = b^5 c^5 ef\\
    z & = acf\\
    w & = bde
  \end{split}
\end{equation}
with the constraint $xy = z^5w^5$. As discussed in
\cite{Franco:2007ii}, the line orientifold in
figure~\ref{fig:orbiconi} acts as:
\begin{equation}
  z \leftrightarrow w, \qquad x \to x, \qquad y \to y
\end{equation}
We can easily read the action on the GLSM fields $a,b,c,d$ from here,
it is given by:
\begin{equation}
  a \leftrightarrow d, \qquad b \leftrightarrow c,\qquad e
  \leftrightarrow f
\end{equation}
Looking to the toric diagram in figure~\ref{fig:Z5coni}, this implies
that the orientifold acts as a reflection of the toric diagram along
the dashed line. We will present further evidence for this action,
from the point of view of mirror symmetry, in
section~\ref{sec:orientifold-mirror}.

Let us now describe how to join the supersymmetry breaking sector
to the visible and dark matter sectors. As in previous sections, we
 model the latter by two copies of $\bC^3/\bZ_3$.  Looking to
figure~\ref{fig:dp0}a, it is clear that we need to perform a
$SL(2,\bZ)$ transformation in order to be able to join the
subsectors. It is most convenient to transform the $\bC^3/\bZ_3$
geometry using the following $SL(2,\bZ)$ action:
\begin{equation}
  T = \begin{pmatrix}
    0 & -1\\ 1 & 1
    \end{pmatrix}
    \label{eq:dp0transform}
\end{equation}
This transformation brings the toric diagram in figure~\ref{fig:dp0}a
into the form shown in figure~\ref{fig:dp0transform}.
\FIGURE{
  \includegraphics[width=0.2\textwidth]{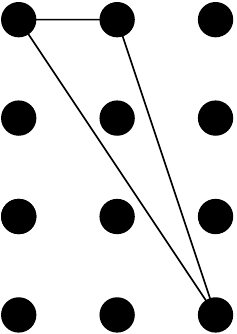}

  \caption{Toric diagram of $\bC^3/\bZ_3$ after
    applying~\eqref{eq:dp0transform}.}

  \label{fig:dp0transform}
}

Now we can easily attach the top of figure~\ref{fig:dp0transform} to
the bottom of \ref{fig:Z5coni}. Due to the action of the orientifold,
we also need to attach a mirror copy on the top. The result of doing
this (including also another copy of $\bC^3/\bZ_3$ for the dark matter
sector) is shown in figure~\ref{fig:dm-improved}.

First, notice that we cannot just join the three sectors and obtain a
well defined toric geometry, since the resulting toric diagram would
not be convex. This is easily remedied by adding the piece in the
lower right of figure~\ref{fig:dm-improved} (and its image on the top
right), which corresponds to a $\bC^3/(\bZ_3\times \bZ_4)$ orbifold of
flat space. When assigning the ranks of the faces of the daughter
dimer models we can choose to put no branes in the daughter diagram
corresponding to this extra orbifold, so this part of the geometry
does not contribute any massless degrees of freedom. Nevertheless it 
modifies the details of the resulting theory, in particular the
couplings to the massive sector fields.  \FIGURE{
  \includegraphics[width=0.6\textwidth]{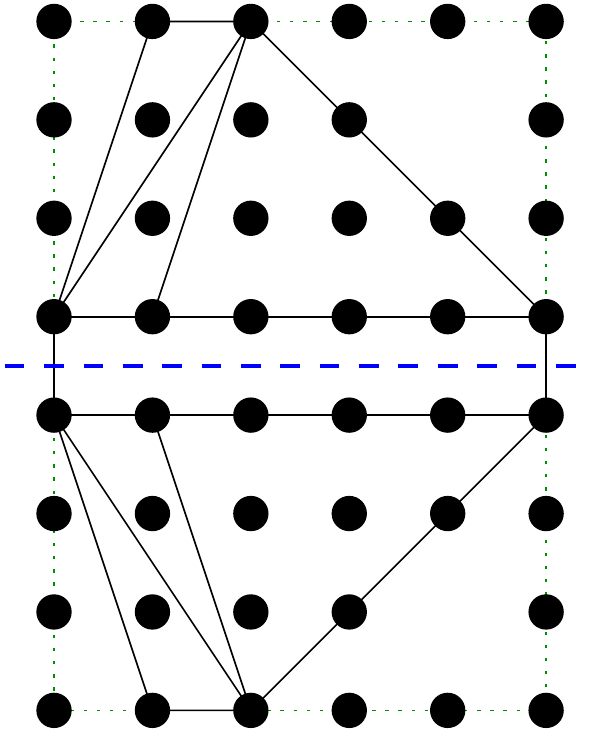}

  \caption{Three sector moose with an improved supersymmetry breaking
    subsector. We have denoted the orientifold point by the blue
    dashed horizontal line. The green dotted rectangle around the
    figure denotes the $\bZ_5\times \bZ_7$ orbifold of the conifold
    which can be used to obtain the gauge theory of our model, as
    explained in the text.}

  \label{fig:dm-improved}
}

Even if the total geometry is complicated, it is actually
straightforward to work out the dimer model description for the joined
singularity without having to go through the joining algorithm. The
basic observation is that the singularity that we get can be easily
embedded into a $\bZ_5\times \bZ_7$ orbifold of the conifold, as shown
by the dotted line in figure~\ref{fig:dm-improved}. Taking orbifolds
of dimer models is very simple, and reduces to increasing the size of
the unit cell \cite{Hanany:2005ve}. Going from the orbifold of the
conifold to our desired geometry involves removing a
$\bC^2/\bZ_3\times \bC$ on the top left (and its image), and a
$\bC^3/(\bZ_3\times\bZ_3)$ piece on the top right (and its
image). This is  done in the dimer using the resolution
algorithm described in section~\ref{sec:partial-resolution}, so we
skip the detailed discussion.

\section{Mirror description of the small resolution}
\label{sec:mirror}

So far we have focused on the type IIB picture, reviewing the set of
rules that describe the small resolution in terms of a Higgsing of the
field theory living on the branes located at the singularity.

Our type IIB system has a well known mirror description
\cite{Hori:2000kt,Hori:2000ck}. In fact, as was shown in
\cite{Feng:2005gw}, the dimer model construction admits a very natural
interpretation in the mirror picture. The effect of the small resolution on
the field theory reviewed above can also be motivated from the mirror
\cite{GarciaEtxebarria:2006aq}. In this section we
describe the mirror type IIA intersecting brane system, and in
particular discuss in some detail how to obtain the complex structure
deformation that is mirror to the small resolution we want to perform in type
IIB.

The dimer model is most naturally seen as a hybrid description of the
system in the sense that it encodes the relevant information for both
sides of the mirror. Any statement that we make in terms of the dimer
model can be easily read in the mirror type IIA language, and allows us to
explicitly implement any of our modular constructions also in the
class of intersecting brane models mirror to branes at toric
singularities.\footnote{See for example \cite{Uranga:2002pg} for an
  example of some interesting model building possibilities using type
  IIA geometries similar to the ones that we construct.}

\subsection{The mirror manifold}

Let us review in this section the construction of the mirror manifold,
referring the reader to the original references
\cite{Hori:2000kt,Hori:2000ck,Feng:2005gw} for more details.

Take any toric Calabi-Yau variety, described by a polytope in the
two-dimensional square lattice. For illustrative purposes we choose
the complex cone over $dP_0$, shown in figure~\ref{fig:dp0toric}. It
is useful to distinguish between \emph{external} and \emph{internal}
points in the toric diagram. In the case of $dP_0$ we have the
external points $(1,0),(0,1),(2,2)$, and a single internal point at
$(1,1)$.\footnote{Toric diagrams without internal points require extra
  care \cite{Feng:2005gw,Franco:2007ii}. Except for one of our
  examples in section~\ref{sec:amoeba}, we work with toric
  singularities \emph{with} internal points.}

\FIGURE{
  \includegraphics[width=0.4\textwidth]{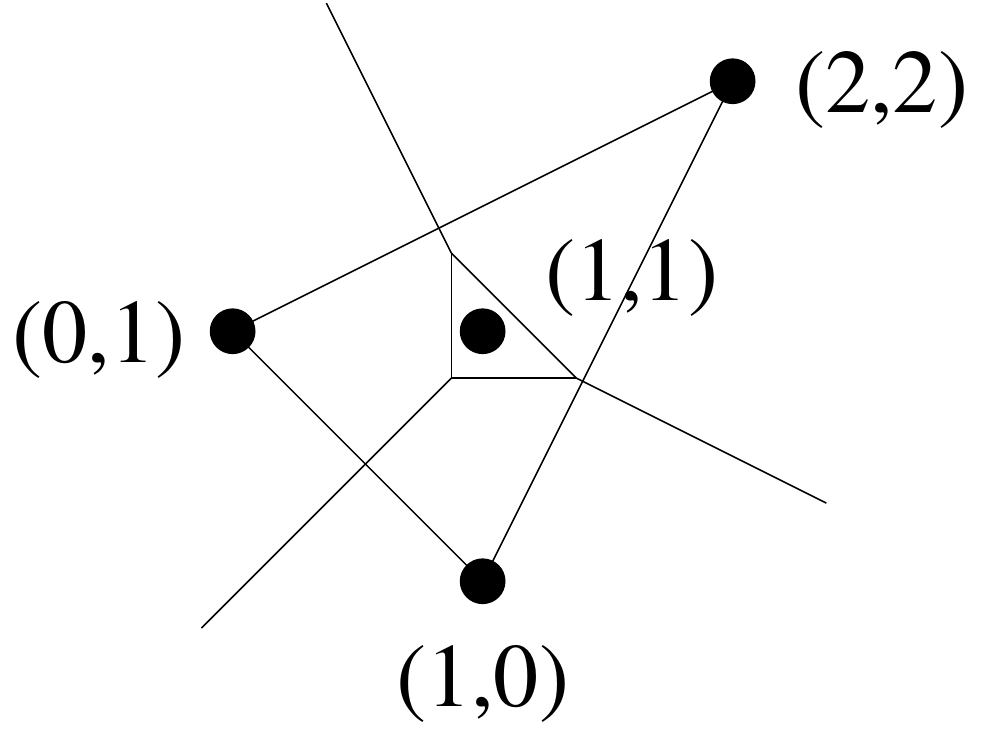}

  \caption{Toric diagram and web diagram for the $\bC^3/\bZ_3$
    orbifold, also known as the complex cone over $dP_0$. The labels
    denote the positions in the square lattice of the nodes in the
    toric diagram.}

  \label{fig:dp0toric}
}

The type IIA mirror manifold is given by a double fibration over the
complex plane $\bC$. Let us denote the coordinate in the complex plane
as $W$. The equations defining the fibration are:
\begin{align}
  uv &= W\label{eq:uv-fiber}\\
  P(z,w) & =W\label{eq:riemann-fiber}
\end{align}
where $u,v\in \bC$ and $z,w\in \bC^*$.  Here, $P(z,w)$, the
\emph{Newton polynomial}, is given by the
equation
\begin{equation}
  P(z,w) = \sum_{(p,q)\in D} c_{(p,q)}z^p w^q
\end{equation}
where the sum is over the positions of the points in the lattice
belonging to the toric diagram (including possible internal points),
and $c_{(p,q)}$ are arbitrary complex numbers parameterizing the
complex structure moduli space of the mirror manifold. For $dP_0$ we
get the following Newton polynomial:
\begin{equation}
  \label{eq:dP0-Newton}
  P(z,w) = c_{(1,0)} z + c_{(0,1)} w + c_{(1,1)} zw + c_{(2,2)} z^2 w^2.
\end{equation}
Notice that there is an ambiguity in the definition of the
polynomial, since we have to choose a particular origin of the
lattice, and a $SL(2,\bZ)$ frame for the embedding. These choices
give rise to isomorphic geometries, since they both arise from the
underlying $SL(3,\bZ)$ ambiguity of the toric description of the
singularity. Acting with this $SL(3,\bZ)$ changes the coordinates
which we use to describe the GLSM formulation of the toric singularity,
but it does not change the toric space itself. 

Equation~\eqref{eq:uv-fiber} describes a $\bC^*$ that degenerates over
$W=0$ into two intersecting complex planes, while the most interesting part
of the geometry comes from~\eqref{eq:riemann-fiber}. This
equation describes a Riemann surface, $\Sigma$, of genus equal to the
number of internal points in the toric diagram, fibered over the
complex plane $W$. $\Sigma$ is actually non-compact, having one
puncture for each external leg of the web diagram. Note that this
surface admits a nice description as a thickening of the web diagram,
an observation that we will use extensively in
section~\ref{sec:amoeba}. For an
illustration of the thickening in the case of $\bC^3/\bZ_3$, see figure~\ref{fig:dp0riemann}.

\FIGURE{
  \includegraphics[width=0.4\textwidth]{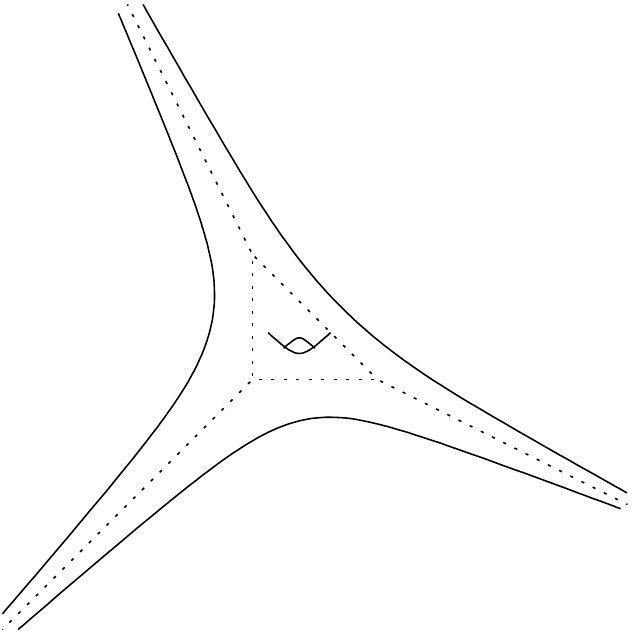}

  \caption{Thickening of the $\bC^3/\bZ_3$ web diagram, giving the
    mirror Riemann genus 1 surface $\Sigma$. We have also superposed,
    using dotted lines, the skeleton web diagram.}

  \label{fig:dp0riemann}
}

The Riemann surface $\Sigma$ degenerates at $n$ critical points,
$W=W_*$, where $n$ is the number of triangles in the toric
diagram. For example, for $\bC^3/\bZ_3$ we have that the toric diagram
is made of 3 elementary triangles, so we expect~\eqref{eq:dP0-Newton}
to degenerate at 3 points. It is a simple matter to check that
$\partial_z P(z,w) = \partial_w P(z,w) = P(z,w)-W= 0$ does indeed have
three different solutions for the three unknowns $(z,w,W)$. Notice
that all the information about where in the base $\Sigma$ degenerates
is encoded in the $c_{(p,q)}$ coefficients, so giving these is
sufficient to specify the geometry completely. Because of this fact,
in the discussion given below we will focus only on $\Sigma$ itself,
with the understanding that this information is enough to define a
mirror Calabi-Yau manifold via the construction given above.

Before proceeding with mapping K\"ahler deformations to complex
structure deformations in the mirror manifold, for completeness let us
say a few words about the mirror description of the (fractional) D3
branes. As discussed above, there are $n$ points $W_*$ on the base
where some one-cycle in $\Sigma$ degenerates. We also have that the
$S^1$ circle in the $\bC^*$ fiber degenerates over $W=0$. If we
connect $W=0$ to any of the $n$ points $W_*$ by a segment in the $W$
plane, and consider the total space of the fibration given by the two
$S^1$ circles degenerating at the boundary of the segment, we obtain a
$S^3$. Since we have $n$ critical points $W_*$, we have $n$ such
three-spheres. The mirror of the $n$ different fractional branes in
the original geometry\footnote{Remember that $n$ is the number of
  triangles in the toric diagram, which coincides with the number of
  nodes of the quiver.} are the different D6 branes wrapping these
$S^3$ cycles. The dimer model then arises from the intersection of the
$S^3$ cycles over $W=0$, and in particular can be understood purely
from intersections of one-cycles over $\Sigma$. We refer the reader to
\cite{Feng:2005gw} for further details on how this construction works,
and how to understand many of the features of the dimer model from it.

\subsection{The amoeba map and small resolutions}
\label{sec:amoeba}

Up to now we have described the mirror at a general point in complex
structure moduli space, without yet specifying the values of the
$c_{(p,q)}$ coefficients in the Newton polynomial $P(z,w)$. Recall
that we are interested in K\"ahler deformations that partially smooth
out the singularity, meaning that we blow up some internal two- (or
four-) cycles, leaving  separated daughter
singularities. In general, mapping K\"ahler deformations to complex
structure deformations of the mirror is a complicated
problem. Fortunately, as we have seen there is a deep relation between
$\Sigma$ and the web diagram of our singularity, and this allows us to
see which complex deformations do the job of splitting the Riemann
surface into two daughter Riemann surfaces with the right properties.

A tool that we  find particularly useful is the \emph{amoeba map}
\cite{mikhalkin-2000-2,mikhalkin-2001,rullgard,kenyon-2003,mikhalkin-2004},
defined as the image in $\bR^2$ of the points of $\Sigma$ (defined as
the locus $P(z,w)=0$) under:
\begin{equation}
  (z,w) \to (\log|z|, \log|w|)
\end{equation}
We illustrate the result of applying this map to the complex cone over
$dP_0$ in figure~\ref{fig:dp0amoeba}.

\FIGURE{
  \includegraphics[width=0.4\textwidth]{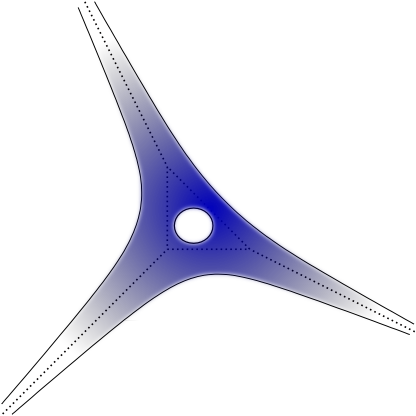}
  
  \caption{Amoeba for the Riemann surface mirror to the $\bC^3/\bZ_3$
    orbifold. We have indicated the web diagram skeleton by the dotted
    line.}

  \label{fig:dp0amoeba}
}

An important property of the amoeba is that its tentacles align with
the external legs of the web diagram (see
figure~\ref{fig:dp0amoeba}). It is useful for us to work through
some examples explicitly, so let us show this for the $dP_0$
singularity.  (For a general proof, see \cite{Feng:2005gw}.)  Let us
take the following expression for the Newton polynomial defining
$\Sigma$:
\begin{equation}
  \label{eq:dp0newton}
  P(z,w) = z + w + d zw + z^2w^2 = 0,
\end{equation}
where we have used the freedom to redefine the coordinates in order to make three
coefficients equal to 1. The K\"ahler deformation parameter of
$\bC^3/\bZ_3$ is encoded in the value for $d$. Let us focus on the
lower left leg of the web diagram, see figure~\ref{fig:dp0toric}. It is perpendicular to the edge
going from $(1,0)$ to $(0,1)$ in the toric diagram. It is easy to see
that there is such a tentacle of the amoeba: it is given by the family
of solutions $z=-w\to 0$ (where we discarded higher order terms
in~\eqref{eq:dp0newton}). Under the amoeba map, $(z, -z)$ with $z\to
0$ maps to $(t,t)$ with $t\to -\infty$, reproducing the proper spike.

The other two tentacles work in a similar way. Let us focus on the
tentacle going northwest. The corresponding leg in the web diagram is
associated with the edge going from $(0,1)$ to $(2,2)$. In terms of
the Newton polynomial, these are the $w$ and $z^2w^2$ terms. Let us
thus write the Newton polynomial equation in the following way:
\begin{equation}
  P(z,w) = w(1+z^2w) + z + d zw = 0.
\end{equation}
Consider the branch of the solution with $w\to\infty$. In order
to keep the product in the first term finite, we need $z^2 = -1/w\to
0$. Note that the last term $dzw$, although divergent, is subleading
and can be ignored when computing the asymptotic behavior of the
amoeba. We end up with the following curve of solutions: $(z,-1/z^2)$
with $z\to 0$, which maps to $(t,-2t)$ with $t\to-\infty$, matching
the result from the web diagram. Finally, in order to obtain the
south-east tentacle, write the curve as:
\begin{equation}
  P(z,w) = z(1+zw^2) + w + dzw = 0.
\end{equation}
There is a branch of solutions with $z\to\infty$ and $zw^2=-1$, in
other words $(-1/w^2,w)$ with $w\to 0$, mapping to $(2t,-t)$ with
$t\to\infty$.

The previous discussion can be seen as a way of ``undoing mirror
symmetry'', in the sense that the skeleton of the amoeba is given by
the web diagram, and represents faithfully the original toric
geometry. It is a natural question whether the same projection can be
used to map complex structure moduli deformations to K\"ahler moduli
deformations. As we will see in examples below, this is actually
possible. In particular, we will argue that it is possible to identify
which complex structure deformations are dual to which K\"ahler deformations
splitting the original singularity into two smaller singularities.

Before explaining in some selected examples how this works, let us
mention a couple of limitations of the method we will present
momentarily. First of all, we are computing properties of the
asymptotics of the amoeba, so we do not see any K\"ahler
deformation that leaves these invariant. Consider the
$\bC^3/\bZ_3$ geometry we have been using to illustrate our
discussion. As we saw, the $d$ parameter is subleading, and does not
influence the asymptotics. This is also easy to see from the web
diagram: we can blow up the collapsed 4-cycle keeping the external
legs fixed. Luckily, the kinds of the small resolutions that we 
deal with in this paper are not of this kind: when we partially blow
up the singularity the external legs get displaced. This is in fact
the basic idea of our method: we can compute this displacement easily
and reliably looking to the asymptotics. Whichever complex deformation
displaces the tentacles of the amoeba in the right way gets identified
with the mirror of the K\"ahler blow-up mode.

Secondly, when computing the behavior of the external tentacles the
coefficients of internal points are often subleading, and hard to
determine looking only to asymptotics. In complicated examples we just
assume that we know their correct value so they reproduce the right
mirror for the singular daughter geometries. This is not a big problem
for us, since we are mostly interested in giving an explicit
description of how to deform the mirror so we end up with various
decoupled (at the massless level) sectors. Knowing which moduli we
have to send to 0 or $\infty$ is enough for understanding this.

With these limitations in mind, let us discuss how we can understand
the geometry of the mirror better in a simple way in a few interesting
examples.

\subsubsection*{The conifold}

We  start by considering our favorite example, the conifold. Its Newton
polynomial equation is:
\begin{equation}
  \label{eq:coni-newton}
  P(z,w) = 1 + z + wz + aw = 0
\end{equation}
where we have defined the variables in such a way that the last first
coefficients are equal to 1. The complexified resolution parameter of
the conifold is then identified with the complex structure modulus
$a$. Let us see how we can see this using the amoeba projection. The
west spike of the amoeba is associated with the $1+aw$ term. We can
satisfy~\eqref{eq:coni-newton} by setting $z\to 0$, $w=-1/a$. Under
the amoeba map, this maps to the line $(t,-\log|a|)$ with
$t\to-\infty$. It is easy to work out the locations of the other
tentacles in a similar way, the result is:
\begin{align}
  \text{South:} & \qquad (0,t), \quad t\to-\infty\\
  \text{West:} & \qquad (t,-\log|a|), \quad t\to-\infty\\
  \text{North:} & \qquad (log|a|,t), \quad t\to\infty\\
  \text{East:} & \qquad (t,0), \quad t\to\infty
\end{align}
Note that once we send $\log|a|\to-\infty$ this exactly reproduces the
web diagram description (given in figure~\ref{fig:coniresolvdimer}a)
of the resolution of the conifold into two copies of flat space. This
tells us that the right limit of~\eqref{eq:coni-newton} to reproduce
the splitting is $a\to 0$, and more generally that we can identify the
resolution parameter $v$ of the conifold with $-\log|a|$.

This is in fact also natural from the point of view of the toric
diagram. Sending $a\to 0$ takes~\eqref{eq:coni-newton} into
\begin{equation}
  P(z,w) = 1+z+wz = 0
\end{equation}
which reproduces the equation obtained from the toric diagram of flat
space. This result is general, at least for the examples that we will
be dealing with, and can be stated very simply as follows: \emph{in
  order to obtain the mirror of the resolved geometry, send to 0 the
  coefficients of the Newton polynomial located purely on one side of
  the resolution.}

\subsubsection*{The complex cone over $\bF_0$}

Let us check this rule in a couple of further examples. The first
example we want to discuss is the complex cone over $\bF_0=\bP^1\times
\bP^1$. The toric and web diagrams are shown in
figure~\ref{fig:F0toric}.

\FIGURE{
  
\ifpdf
  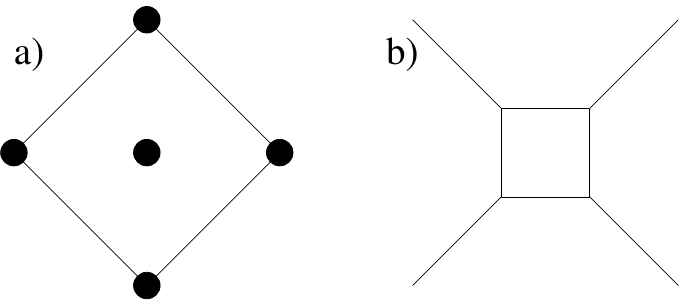
  \else
  \input{F0toric.pstex_t}
\fi

  \caption{a) Toric diagram for the complex cone over $\bF_0$. We have
    indicated the corresponding terms in the Newton
    polynomial~\eqref{eq:F0-newton}. b) Web diagram. The $\alpha$ and
    $\beta$ parameters correspond to the size of the two independent
    blow-up modes of $\bF_0$.}

  \label{fig:F0toric}
}

The Newton polynomial equation defining the geometry can be taken to
be:
\begin{equation}
  \label{eq:F0-newton}
  P(z,w) = z + w + zw + azw^2 + bz^2 w
\end{equation}
We can determine the slope and displacement of the external legs
easily using the same methods as above. For example, in order to
compute the displacement of the south-east tentacle, we need to center
on the $z+bz^2w$ term in the Newton polynomial. Rewriting
~\eqref{eq:F0-newton} as:
\begin{equation}
  P(z,w) = z(1+bzw) + w + zw + azw^2
\end{equation}
we obtain a tentacle of the amoeba by sending $z\to\infty$ while
keeping $w=-1/bz$. That is, we obtain the line given by
$(t,-t-\log|b|)$ with $t\to\infty$. We proceed similarly for the other
external legs, obtaining:
\begin{align}
  \text{South-West:}&\qquad (t,t),\quad t\to\infty\\
  \text{South-East:}&\qquad (t-\log|b|,-t) \quad t\to\infty\\
  \text{North-East:}&\qquad (t-\log|b|,t-\log|a|)\quad t\to\infty\\
  \text{North-West:}&\qquad (-t,t-\log|a|)\qquad t\to\infty
\end{align}
It is easy to see that this is precisely the structure of external
legs in the resolved web diagram shown in figure~\ref{fig:F0toric}b,
once we identify $\alpha=-\log|a|$ and $\beta=-\log|b|$. This also
matches beautifully with the rule described above: if we wanted to
resolve the geometry into a couple of $\bC^2/\bZ_2\times \bC$
singularities we tune the K\"ahler moduli to be
$\alpha=0,\beta\to\infty$. This is precisely $a=1,b=0$, which removes
the leftmost node in the toric diagram in figure~\ref{fig:F0toric}a
from the Newton polynomial~\eqref{eq:F0-newton}.

\subsubsection*{The $X^{(3,1)}$ geometry}

Let us discuss another example, given by the mirror of the $X^{(3,1)}$
geometry we used in section~\ref{sec:dark-mssm} for modelling the MSSM
and susy breaking sectors in our three sector dark matter model. We
reproduce in figure~\ref{fig:X31-newton} the relevant toric and web
diagrams.

\FIGURE{
\ifpdf
  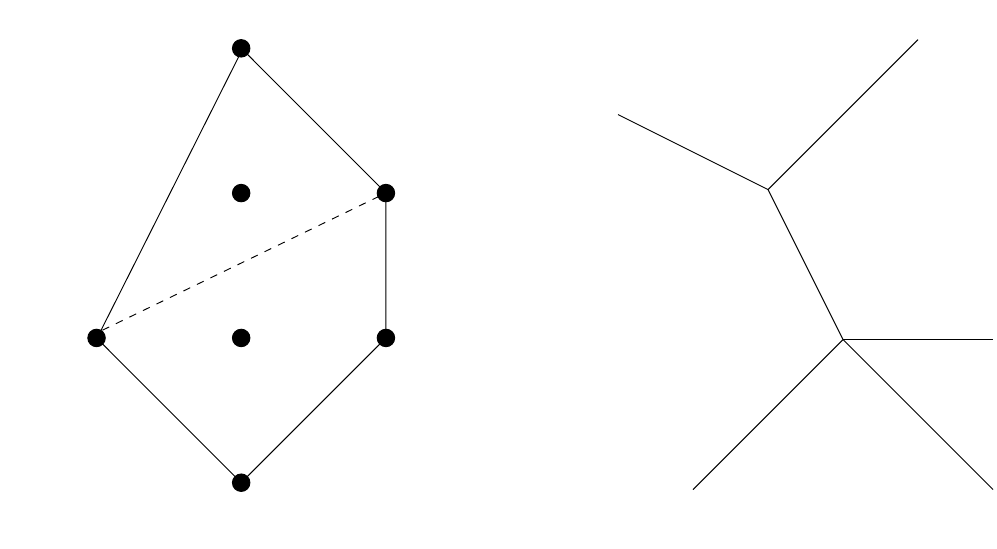
  \else
  \input{X31-newton.pstex_t}
\fi

  \caption{a) Toric diagram for the MSSM and susy breaking sectors in
    section~\ref{sec:dark-mssm}. We have indicated the desired
    resolution mode by the dashed lines. We have also indicated which
    terms in~\eqref{eq:X31-newton} correspond to which vertex. b)
    Corresponding web diagram. Note that we have not resolved the
    internal four-cycles in order to make better contact with the
    text. We have denoted the size of the resolved two-cycle by
    $\alpha$.}

  \label{fig:X31-newton}
}

We have a Newton polynomial given by:
\begin{equation}
  \label{eq:X31-newton}
  P(z,w) = w + azw^3 + w^2z^2 + wz^2 + z = 0
\end{equation}
where we have omitted the terms corresponding to the internal points,
since they do not modify the asymptotics. As described above, we
expect that setting $a\to 0$  splits the Riemann surface into the
two desired daughter Riemann surfaces. We obtain the following
asymptotics for the tentacles of the amoeba:
\begin{align*}
  \text{North-West:}&\quad (-2t-\log|a|, t),\quad t\to\infty\\
  \text{North-East:}&\quad (t+\log|a|, t), \quad t\to\infty\\
  \text{East:}&\quad (0,t),\quad t\to\infty\\
  \text{South-East:}&\quad (t,-t),\quad t\to\infty\\
  \text{South-West:}&\quad (-t,-t),\quad t\to\infty
\end{align*}
It is easy to check that these lines reproduce the structure of the
resolved web diagram for $X^{(3,1)}$ shown in
figure~\ref{fig:X31-newton}b, once we set $\alpha=-\log{|a|}/3$. In
particular, the NW and NE external legs intersect over
$(-\alpha,2\alpha)$ in the plane where the amoeba is defined. The rest
of the legs intersect over $(0,0)$.

Let us discuss an important point that we have ignored until now, and
which will become useful in a moment. Notice that in doing the blow up
both daughter sub-sectors enter on an equal footing, while our
construction seems to depend strongly on which sub-sector contains the
point that we want to ``remove'' by sending $a\to 0$.  However, there
is a way of rewriting things in such a way that the same resolution
can be described by sending the coefficients of the points on the
other side of the resolution to zero.  Let us describe how this works
in our $X^{(3,1)}$ example. The key observation is that the position
of the origin of the amoeba plane is conventional, so in particular we
should be able to scale our $z,w$ variables in order to put the NW-NE
intersection at the origin. The relevant scaling is given by $\tilde
z=z/\tilde a, \tilde w=w\tilde a^2$ with $\tilde a^3=a$. In terms of
these variables the Newton polynomial equation becomes:
\begin{equation}
  P(\tilde z, \tilde w) = \frac{1}{\tilde a^2}(\tilde z\tilde w^3 +
  \tilde w + \tilde w^2 \tilde z^2) + \tilde a \tilde z + \tilde
  w\tilde z^2 = 0.
\end{equation}
Multiplying the whole equation by $\tilde a^2$ we get:
\begin{equation}
  \tilde P(\tilde z, \tilde w) = \tilde z\tilde w^3 +
  \tilde w + \tilde w^2 \tilde z^2 + \tilde a^3 \tilde z + \tilde
  a^2w\tilde z^2 = 0.
\end{equation}
Notice how the resulting polynomial now agrees with the general
prescription when applied to the other side of the resolution: setting
$\tilde a\to 0$ removes the terms in the Newton polynomial
corresponding to that side of the resolution, and indeed resolves the
singularity as desired. In this case the powers of $\tilde a$ are
non-trivial, and are determined by imposing that the external legs
corresponding to the $dP_1$ part of the geometry intersect at a single
point.

\subsubsection*{The dark matter model}

\FIGURE{
  
\ifpdf
  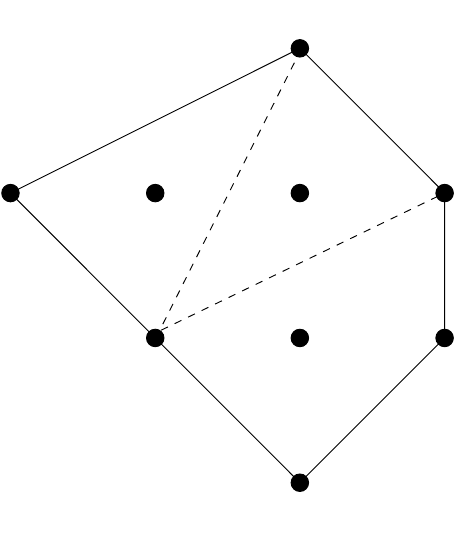
  \else
  \input{dm-newton.pstex_t}
\fi

  \caption{Toric diagram for the dark matter model in
    section~\ref{sec:dark-mssm}. We have indicated the desired
    resolutions by dashed lines, and we have associated monomials
    in~\eqref{eq:dm-newton} to their corresponding edges.}

  \label{fig:dm-newton}
}

Just for completeness, let us include here the result for the mirror
of our three-sector dark matter model from
section~\ref{sec:dark-mssm}. The relevant toric diagram is shown in
figure~\ref{fig:dm-newton}. The Newton polynomial equation (omitting
internal points, as usual) is given by:
\begin{equation}
  \label{eq:dm-newton}
  P(z, w) =  bw^2 + z^2w^3 + zw + w^2z^3 +  a^3z^2 + a^2w z^3= 0.
\end{equation}
The discussion proceeds just as in the examples given above, and we
obtain the factorized three-sector geometry by sending $a\to 0$, $b\to
0$.

\medskip

As we have seen in the previous examples, looking to the displacement
of the tentacles via the amoeba map gives a simple way of determining
how to deform the Newton polynomial in order to split the system into
decoupled sub-sectors. Our construction gives a simple and very
intuitive procedure for determining how to deform the Riemann surface.

\subsection{Orientifolds and algae}
\label{sec:orientifold-mirror}

In section~\ref{sec:improved-breaking} we discussed the geometric
action of the orientifold in the toric diagram, or equivalently in the
web diagram, and argued from the defining GLSM
that the proper geometric action is a reflection along a horizontal
line. Let us shortly describe how the mirror picture supports this
identification.

We described in the previous section how the amoeba map gives a way to
map the behavior of the Riemann surface to the mirror description in
terms of the web diagram. There is a complementary projection of the
Riemann surface called the \emph{alga projection},
defined by:
\begin{equation}
  (z,w) = (|z|e^{i\alpha},|w|e^{i\beta}) \to (\alpha,\beta).
\end{equation}
This projection defines a region on a two torus that turns out to be a
thickening of the dimer model \cite{Feng:2005gw}. Let us remark that
the dimer model that we obtain is naturally ``aligned'' with the web
diagram, in the sense that if a leg of the web diagram goes in a
$(p,q)$ direction in the amoeba plane, then the corresponding zig-zag
path wraps a $(p,q)$ cycle in the dimer model $T^2$.

As described in \cite{Franco:2007ii}, the above observation can be
used to understand the behavior of the mirror surface under the
orientifold action. The basic idea is that the phases of $z,w$ must
transform as the coordinates of the dimer $T^2$. Since we know the
action of the orientifold on the dimer we can infer its action on the
phases of $z,w$. Combined with the fact that the orientifold action on
the mirror must be anti-holomorphic, we can determine how $z,w$
transform under the orientifold.

Consider the configuration studied in
section~\ref{sec:improved-breaking}. The line orientifold reflects the
dimer horizontally, while leaving it invariant vertically. An
anti-holomorphic involution that does this is \cite{Franco:2007ii}:
\begin{align}
  \label{eq:mirror-involution}
  \begin{split}
    z & \to \ov z\\
    w & \to \frac{1}{\ov w}
  \end{split}
\end{align}
Notice that the phase of $z$ changes sign, while the phase of $w$ does
not, according to the fact that the reflection of the dimer model is
along the horizontal direction.

It is easy to see the action that~\eqref{eq:mirror-involution} induces
on the amoeba. It is given by:
\begin{equation}
  (\log|z|,\log|w|) = (s,t) \to (s,-t)
\end{equation}
The horizontal reflection has turned into a vertical one. Reading the
action on the web diagram from the action on the amoeba, this exactly
reproduces the geometric action found in
section~\ref{sec:improved-breaking}.

\section{Conclusions and open questions}
\label{sec:conclusions}

We have described a general procedure for joining arbitrary theories
coming from branes at toric singularities into a consistent local
embedding which couples them via massive mediators. We have applied
this technique for building a three sector model of dark matter
consisting of the standard model, a dark matter sector, and a
supersymmetry breaking sector, all interacting via massive mediators
of a tunable scale. We have also constructed some alternative models
in order to illustrate the flexibility of our constructions. In
particular, note that some of our examples can be naturally adapted
(by forgetting about the dark matter sector) to become GMSB scenarios
with a nicely behaved susy breaking sector. Finally, we have described
the mirror picture of our construction, giving us a procedure for
joining a large class of intersecting brane configurations
(specifically, those mirror to branes at toric singularities).

\medskip

There are various interesting directions in which our results can be
extended. The most obvious one is to generalize our construction to
models of branes at non-toric singularities. In particular, the case
of branes at singular non-toric del Pezzos can be analyzed using
exceptional collections \cite{Cachazo:2001sg,Wijnholt:2002qz}. It
would be interesting to develop techniques to analyze these cases in a
way similar to the one we have developed in this paper.\footnote{The
  first four complex cones over del Pezzo are actually toric, and can
  be understood both from the point of view of exceptional collections
  and dimer models. Furthermore, for these cases the map between the
  two approaches is well understood by now \cite{Hanany:2006nm}.}
Being able to handle higher del Pezzo singularities would allow for
some very interesting model building possibilities
\cite{Verlinde:2005jr,Buican:2006sn}.

\medskip 

Another related line of research concerns the relation to recent model
building work based on local F-theory constructions (see
\cite{Donagi:2008ca,Beasley:2008dc,Hayashi:2008ba,Beasley:2008kw,Donagi:2008kj}
for some of the original papers). Most of the subsectors in this paper
are based on blown down del Pezzo surfaces. The same kind of geometry,
but blown up and having a different choice of wrapped brane, can be
used to construct semi-realistic models of the MSSM GUTs in
F-theory. It would be interesting to construct ``hybrid'' models,
having the visible sector coming from a D7 wrapping a large blown-up
del Pezzo, while having branes at singularities nearby providing for
susy breaking and interacting dark matter. In fact some of these
hybrid models have already been studied in the literature
\cite{Diaconescu:2005pc}.

\medskip

The models that we have constructed are based on small resolutions, and are
thus controlled by K\"ahler moduli. In our local constructions these
are flat directions. We have assumed that we can tune the K\"ahler
moduli as we wish, but in any realistic setting one must construct a
detailed moduli stabilization scenario. This  depends on global
features of our model, and is thus beyond the reach of the tools
discussed in this paper. Related to this last point, we expect the
global embedding of our construction to introduce interesting
constraints in the local physics. It would thus be useful to have a
better understanding of possible global completions of our models.

There is an alternative to our construction which improves on this
moduli stabilization problem, at the cost of making the models less
calculable. The idea is to trigger geometric transitions by putting
confining D-branes on the singularity. Once the branes confine, we end
up with daughter singularities separated by three cycles stabilized by
the dual flux. Similar techniques to the ones presented in this paper
apply to this scenario
\cite{Franco:2005zu,GarciaEtxebarria:2006aq,GarciaEtxebarria:2006rw}. The
problem with this setup is that one losses calculability of the
mediator sector, which is now determined by the spectrum of states in
the confining theory. This is a well defined question, but the answer
is very hard to determine with available tools.

As we discussed in section~\ref{sec:mirror}, our analysis carries over
easily to the context of intersecting branes in type IIA. Under mirror
symmetry, the resolutions that we have constructed in type IIB map to complex
structure deformations of the type IIA background. In type IIA it is
possible to stabilize complex structure moduli by the introduction of
$H_3$ and geometric flux
\cite{Derendinger:2004jn,Kachru:2004jr,Villadoro:2005cu,Derendinger:2005ph},
so the problem of moduli stabilization may in principle be
ameliorated. Unfortunately, studying in detail the backreaction of the
flux on the type IIA geometry is an involved problem, and thus it is
hard to construct explicit models with all moduli stabilized.

\medskip

Finally, in this paper we have focused mainly on developing tools, and
understanding the construction better via some toy models. It would be
interesting to construct more realistic models of interacting dark
matter and/or gauge mediated supersymmetry breaking using these tools,
and analyze their phenomenology in more detail.

\acknowledgments

We are happy to acknowledge interesting and fruitful discussions with
Mirjam Cveti{\v c}, Angel Uranga and Tomer Volansky. We would also
like to thank Antonio Amariti, Luciano Girardello and Alberto Mariotti
for informing us of their research on a related topic
\cite{Amariti:2009tu}. V.B. thanks the Yukawa Institute for
Theoretical Physics for hospitality while this work was being
completed. P.B. thanks the organizers of the workshop ``New
Perspectives in String Theory'' at the Galileo Galilei Institute for
Theoretical Physics, Florence, where early stages of this work was
carried out. I.G.-E. thanks the CERN Theory Division for hospitality,
and Nao Hasegawa for kind support and constant encouragement.  The
work of P.B. is supported by the NSF CAREER grant PHY-0645686 and by
the University of New Hampshire through its Faculty Scholars Award
Program. I.G.-E.'s work is supported by the High Energy Physics
Research Grant DE-FG05-95ER40893-A020.

\bibliographystyle{JHEP}
\bibliography{refs}

\end{document}

%% file: coniresolv.pdf_t
\begin{picture}(0,0)%
\includegraphics{coniresolv.pdf}%
\end{picture}%
\setlength{\unitlength}{3315sp}%
\begingroup\makeatletter\ifx\SetFigFontNFSS\undefined%
\gdef\SetFigFontNFSS#1#2#3#4#5{%
  \reset@font\fontsize{#1}{#2pt}%
  \fontfamily{#3}\fontseries{#4}\fontshape{#5}%
  \selectfont}%
\fi\endgroup%
\begin{picture}(3730,1374)(3271,-2998)
\put(6931,-2806){\makebox(0,0)[lb]{\smash{{\SetFigFontNFSS{10}{12.0}{\rmdefault}{\mddefault}{\updefault}{\color[rgb]{0,0,0}$z_4$}%
}}}}
\put(6211,-1861){\makebox(0,0)[lb]{\smash{{\SetFigFontNFSS{10}{12.0}{\rmdefault}{\mddefault}{\updefault}{\color[rgb]{0,0,0}$z_3$}%
}}}}
\put(6886,-1861){\makebox(0,0)[lb]{\smash{{\SetFigFontNFSS{10}{12.0}{\rmdefault}{\mddefault}{\updefault}{\color[rgb]{0,0,0}$z_1$}%
}}}}
\put(6211,-2806){\makebox(0,0)[lb]{\smash{{\SetFigFontNFSS{10}{12.0}{\rmdefault}{\mddefault}{\updefault}{\color[rgb]{0,0,0}$z_2$}%
}}}}
\put(5761,-1771){\makebox(0,0)[lb]{\smash{{\SetFigFontNFSS{10}{12.0}{\rmdefault}{\mddefault}{\updefault}{\color[rgb]{0,0,0}b)}%
}}}}
\put(3286,-1771){\makebox(0,0)[lb]{\smash{{\SetFigFontNFSS{10}{12.0}{\rmdefault}{\mddefault}{\updefault}{\color[rgb]{0,0,0}a)}%
}}}}
\put(4456,-2716){\makebox(0,0)[rb]{\smash{{\SetFigFontNFSS{10}{12.0}{\rmdefault}{\mddefault}{\updefault}{\color[rgb]{0,0,0}$z_2=0$}%
}}}}
\put(4231,-1951){\makebox(0,0)[rb]{\smash{{\SetFigFontNFSS{10}{12.0}{\rmdefault}{\mddefault}{\updefault}{\color[rgb]{0,0,0}$z_3=0$}%
}}}}
\put(4636,-2041){\makebox(0,0)[lb]{\smash{{\SetFigFontNFSS{10}{12.0}{\rmdefault}{\mddefault}{\updefault}{\color[rgb]{0,0,0}$z_1=0$}%
}}}}
\put(4816,-2761){\makebox(0,0)[lb]{\smash{{\SetFigFontNFSS{10}{12.0}{\rmdefault}{\mddefault}{\updefault}{\color[rgb]{0,0,0}$z_4=0$}%
}}}}
\put(4591,-2311){\rotatebox{315.0}{\makebox(0,0)[b]{\smash{{\SetFigFontNFSS{10}{12.0}{\rmdefault}{\mddefault}{\updefault}{\color[rgb]{0,0,0}$v$}%
}}}}}
\end{picture}%

%% file: coniquiverdimer.pdf_t
\begin{picture}(0,0)%
\includegraphics{coniquiverdimer.pdf}%
\end{picture}%
\setlength{\unitlength}{2901sp}%
\begingroup\makeatletter\ifx\SetFigFontNFSS\undefined%
\gdef\SetFigFontNFSS#1#2#3#4#5{%
  \reset@font\fontsize{#1}{#2pt}%
  \fontfamily{#3}\fontseries{#4}\fontshape{#5}%
  \selectfont}%
\fi\endgroup%
\begin{picture}(8925,3066)(2236,-4696)
\put(6301,-3211){\makebox(0,0)[lb]{\smash{{\SetFigFontNFSS{14}{16.8}{\rmdefault}{\mddefault}{\updefault}{\color[rgb]{0,0,0}$U(N)_2$}%
}}}}
\put(2251,-2311){\makebox(0,0)[lb]{\smash{{\SetFigFontNFSS{20}{24.0}{\rmdefault}{\mddefault}{\updefault}{\color[rgb]{0,0,0}a)}%
}}}}
\put(7201,-2311){\makebox(0,0)[lb]{\smash{{\SetFigFontNFSS{20}{24.0}{\rmdefault}{\mddefault}{\updefault}{\color[rgb]{0,0,0}b)}%
}}}}
\put(3601,-3211){\makebox(0,0)[rb]{\smash{{\SetFigFontNFSS{14}{16.8}{\rmdefault}{\mddefault}{\updefault}{\color[rgb]{0,0,0}$U(N)_1$}%
}}}}
\put(4951,-3886){\makebox(0,0)[b]{\smash{{\SetFigFontNFSS{11}{13.2}{\rmdefault}{\mddefault}{\updefault}{\color[rgb]{0,0,0}$X_{21}^1,X_{21}^2$}%
}}}}
\put(4951,-1861){\makebox(0,0)[b]{\smash{{\SetFigFontNFSS{11}{13.2}{\rmdefault}{\mddefault}{\updefault}{\color[rgb]{0,0,0}$X_{12}^1,X_{12}^2$}%
}}}}
\put(9676,-3211){\makebox(0,0)[b]{\smash{{\SetFigFontNFSS{14}{16.8}{\rmdefault}{\mddefault}{\updefault}{\color[rgb]{0,0,0}1}%
}}}}
\put(10801,-2311){\makebox(0,0)[b]{\smash{{\SetFigFontNFSS{14}{16.8}{\rmdefault}{\mddefault}{\updefault}{\color[rgb]{0,0,0}2}%
}}}}
\put(8551,-2311){\makebox(0,0)[b]{\smash{{\SetFigFontNFSS{14}{16.8}{\rmdefault}{\mddefault}{\updefault}{\color[rgb]{0,0,0}2}%
}}}}
\put(8551,-4336){\makebox(0,0)[b]{\smash{{\SetFigFontNFSS{14}{16.8}{\rmdefault}{\mddefault}{\updefault}{\color[rgb]{0,0,0}2}%
}}}}
\put(10801,-4336){\makebox(0,0)[b]{\smash{{\SetFigFontNFSS{14}{16.8}{\rmdefault}{\mddefault}{\updefault}{\color[rgb]{0,0,0}2}%
}}}}
\put(10261,-2761){\rotatebox{315.0}{\makebox(0,0)[lb]{\smash{{\SetFigFontNFSS{11}{13.2}{\rmdefault}{\mddefault}{\updefault}{\color[rgb]{0,0,0}$X_{21}^2$}%
}}}}}
\put(9271,-2671){\rotatebox{45.0}{\makebox(0,0)[lb]{\smash{{\SetFigFontNFSS{11}{13.2}{\rmdefault}{\mddefault}{\updefault}{\color[rgb]{0,0,0}$X_{12}^1$}%
}}}}}
\put(8821,-4021){\rotatebox{315.0}{\makebox(0,0)[rb]{\smash{{\SetFigFontNFSS{11}{13.2}{\rmdefault}{\mddefault}{\updefault}{\color[rgb]{0,0,0}$X_{21}^1$}%
}}}}}
\put(10261,-3841){\rotatebox{45.0}{\makebox(0,0)[rb]{\smash{{\SetFigFontNFSS{11}{13.2}{\rmdefault}{\mddefault}{\updefault}{\color[rgb]{0,0,0}$X_{12}^2$}%
}}}}}
\end{picture}%

%% file: dp0mssm.pdf_t
\begin{picture}(0,0)%
\includegraphics{dp0mssm.pdf}%
\end{picture}%
\setlength{\unitlength}{2368sp}%
\begingroup\makeatletter\ifx\SetFigFontNFSS\undefined%
\gdef\SetFigFontNFSS#1#2#3#4#5{%
  \reset@font\fontsize{#1}{#2pt}%
  \fontfamily{#3}\fontseries{#4}\fontshape{#5}%
  \selectfont}%
\fi\endgroup%
\begin{picture}(4143,2952)(7411,-4207)
\put(9451,-2836){\makebox(0,0)[lb]{\smash{{\SetFigFontNFSS{11}{13.2}{\rmdefault}{\bfdefault}{\updefault}{\color[rgb]{0,0,0}$U_R$}%
}}}}
\put(8476,-1486){\makebox(0,0)[lb]{\smash{{\SetFigFontNFSS{11}{13.2}{\rmdefault}{\bfdefault}{\updefault}{\color[rgb]{0,0,0}$D_R$}%
}}}}
\put(10126,-1486){\makebox(0,0)[lb]{\smash{{\SetFigFontNFSS{11}{13.2}{\rmdefault}{\bfdefault}{\updefault}{\color[rgb]{0,0,0}$\overline{D_R}$}%
}}}}
\put(8476,-2761){\makebox(0,0)[lb]{\smash{{\SetFigFontNFSS{11}{13.2}{\rmdefault}{\bfdefault}{\updefault}{\color[rgb]{0,0,0}$Q_L$}%
}}}}
\put(9301,-1561){\makebox(0,0)[lb]{\smash{{\SetFigFontNFSS{11}{13.2}{\rmdefault}{\bfdefault}{\updefault}{\color[rgb]{0,0,0}$U(3)$}%
}}}}
\put(10576,-4111){\makebox(0,0)[lb]{\smash{{\SetFigFontNFSS{11}{13.2}{\rmdefault}{\bfdefault}{\updefault}{\color[rgb]{0,0,0}$U(1)$}%
}}}}
\put(9376,-4036){\makebox(0,0)[lb]{\smash{{\SetFigFontNFSS{11}{13.2}{\rmdefault}{\bfdefault}{\updefault}{\color[rgb]{0,0,0}$H_U$}%
}}}}
\put(8326,-4111){\makebox(0,0)[lb]{\smash{{\SetFigFontNFSS{11}{13.2}{\rmdefault}{\bfdefault}{\updefault}{\color[rgb]{0,0,0}$U(2)$}%
}}}}
\put(7576,-2986){\makebox(0,0)[lb]{\smash{{\SetFigFontNFSS{11}{13.2}{\rmdefault}{\bfdefault}{\updefault}{\color[rgb]{0,0,0}$L$}%
}}}}
\put(7426,-3361){\makebox(0,0)[lb]{\smash{{\SetFigFontNFSS{11}{13.2}{\rmdefault}{\bfdefault}{\updefault}{\color[rgb]{0,0,0}$H_D$}%
}}}}
\put(11251,-3286){\makebox(0,0)[lb]{\smash{{\SetFigFontNFSS{11}{13.2}{\rmdefault}{\bfdefault}{\updefault}{\color[rgb]{0,0,0}$E$}%
}}}}
\end{picture}%

%% file: D7-dimer.pdf_t
\begin{picture}(0,0)%
\includegraphics{D7-dimer.pdf}%
\end{picture}%
\setlength{\unitlength}{2486sp}%
\begingroup\makeatletter\ifx\SetFigFontNFSS\undefined%
\gdef\SetFigFontNFSS#1#2#3#4#5{%
  \reset@font\fontsize{#1}{#2pt}%
  \fontfamily{#3}\fontseries{#4}\fontshape{#5}%
  \selectfont}%
\fi\endgroup%
\begin{picture}(3894,2545)(2689,-4742)
\put(3871,-3841){\makebox(0,0)[lb]{\smash{{\SetFigFontNFSS{14}{16.8}{\rmdefault}{\mddefault}{\updefault}{\color[rgb]{0,0,0}$X_{21}$}%
}}}}
\put(4726,-2536){\makebox(0,0)[b]{\smash{{\SetFigFontNFSS{14}{16.8}{\rmdefault}{\mddefault}{\updefault}{\color[rgb]{0,0,0}$SU(N_1)$}%
}}}}
\put(4726,-4606){\makebox(0,0)[b]{\smash{{\SetFigFontNFSS{14}{16.8}{\rmdefault}{\mddefault}{\updefault}{\color[rgb]{0,0,0}$SU(N_2)$}%
}}}}
\end{picture}%

%% file: Z5coni.pdf_t
\begin{picture}(0,0)%
\includegraphics{Z5coni.pdf}%
\end{picture}%
\setlength{\unitlength}{4144sp}%
\begingroup\makeatletter\ifx\SetFigFontNFSS\undefined%
\gdef\SetFigFontNFSS#1#2#3#4#5{%
  \reset@font\fontsize{#1}{#2pt}%
  \fontfamily{#3}\fontseries{#4}\fontshape{#5}%
  \selectfont}%
\fi\endgroup%
\begin{picture}(2744,1246)(2004,-4085)
\put(4501,-2986){\makebox(0,0)[b]{\smash{{\SetFigFontNFSS{12}{14.4}{\rmdefault}{\mddefault}{\updefault}{\color[rgb]{0,0,0}$b$}%
}}}}
\put(2251,-2986){\makebox(0,0)[b]{\smash{{\SetFigFontNFSS{12}{14.4}{\rmdefault}{\mddefault}{\updefault}{\color[rgb]{0,0,0}$d$}%
}}}}
\put(2251,-4021){\makebox(0,0)[b]{\smash{{\SetFigFontNFSS{12}{14.4}{\rmdefault}{\mddefault}{\updefault}{\color[rgb]{0,0,0}$a$}%
}}}}
\put(4501,-4021){\makebox(0,0)[b]{\smash{{\SetFigFontNFSS{12}{14.4}{\rmdefault}{\mddefault}{\updefault}{\color[rgb]{0,0,0}$c$}%
}}}}
\put(2701,-2986){\makebox(0,0)[b]{\smash{{\SetFigFontNFSS{12}{14.4}{\rmdefault}{\mddefault}{\updefault}{\color[rgb]{0,0,0}$e$}%
}}}}
\put(2701,-4021){\makebox(0,0)[b]{\smash{{\SetFigFontNFSS{12}{14.4}{\rmdefault}{\mddefault}{\updefault}{\color[rgb]{0,0,0}$f$}%
}}}}
\end{picture}%

%% file: F0toric.pdf_t
\begin{picture}(0,0)%
\includegraphics{F0toric.pdf}%
\end{picture}%
\setlength{\unitlength}{1865sp}%
\begingroup\makeatletter\ifx\SetFigFontNFSS\undefined%
\gdef\SetFigFontNFSS#1#2#3#4#5{%
  \reset@font\fontsize{#1}{#2pt}%
  \fontfamily{#3}\fontseries{#4}\fontshape{#5}%
  \selectfont}%
\fi\endgroup%
\begin{picture}(6905,3093)(758,-4315)
\put(2251,-3211){\makebox(0,0)[b]{\smash{{\SetFigFontNFSS{11}{13.2}{\rmdefault}{\mddefault}{\updefault}{\color[rgb]{0,0,0}$zw$}%
}}}}
\put(901,-3211){\makebox(0,0)[b]{\smash{{\SetFigFontNFSS{11}{13.2}{\rmdefault}{\mddefault}{\updefault}{\color[rgb]{0,0,0}$w$}%
}}}}
\put(2431,-1501){\makebox(0,0)[lb]{\smash{{\SetFigFontNFSS{11}{13.2}{\rmdefault}{\mddefault}{\updefault}{\color[rgb]{0,0,0}$azw^2$}%
}}}}
\put(2431,-4201){\makebox(0,0)[lb]{\smash{{\SetFigFontNFSS{11}{13.2}{\rmdefault}{\mddefault}{\updefault}{\color[rgb]{0,0,0}$z$}%
}}}}
\put(3421,-3211){\makebox(0,0)[lb]{\smash{{\SetFigFontNFSS{11}{13.2}{\rmdefault}{\mddefault}{\updefault}{\color[rgb]{0,0,0}$bz^2w$}%
}}}}
\put(6841,-2941){\makebox(0,0)[lb]{\smash{{\SetFigFontNFSS{11}{13.2}{\rmdefault}{\mddefault}{\updefault}{\color[rgb]{0,0,0}$\alpha$}%
}}}}
\put(6301,-3661){\makebox(0,0)[b]{\smash{{\SetFigFontNFSS{11}{13.2}{\rmdefault}{\mddefault}{\updefault}{\color[rgb]{0,0,0}$\beta$}%
}}}}
\end{picture}%

%% file: X31-newton.pdf_t
\begin{picture}(0,0)%
\includegraphics{X31-newton.pdf}%
\end{picture}%
\setlength{\unitlength}{2368sp}%
\begingroup\makeatletter\ifx\SetFigFontNFSS\undefined%
\gdef\SetFigFontNFSS#1#2#3#4#5{%
  \reset@font\fontsize{#1}{#2pt}%
  \fontfamily{#3}\fontseries{#4}\fontshape{#5}%
  \selectfont}%
\fi\endgroup%
\begin{picture}(7957,4377)(-7344,-2632)
\put(-5399,1514){\makebox(0,0)[b]{\smash{{\SetFigFontNFSS{12}{14.4}{\rmdefault}{\mddefault}{\updefault}{\color[rgb]{0,0,0}$azw^3$}%
}}}}
\put(-5399,389){\makebox(0,0)[b]{\smash{{\SetFigFontNFSS{12}{14.4}{\rmdefault}{\mddefault}{\updefault}{\color[rgb]{0,0,0}$zw^2$}%
}}}}
\put(-5399,-2536){\makebox(0,0)[b]{\smash{{\SetFigFontNFSS{12}{14.4}{\rmdefault}{\mddefault}{\updefault}{\color[rgb]{0,0,0}$z$}%
}}}}
\put(-4199,389){\makebox(0,0)[lb]{\smash{{\SetFigFontNFSS{12}{14.4}{\rmdefault}{\mddefault}{\updefault}{\color[rgb]{0,0,0}$w^2z^2$}%
}}}}
\put(-4199,-1336){\makebox(0,0)[lb]{\smash{{\SetFigFontNFSS{12}{14.4}{\rmdefault}{\mddefault}{\updefault}{\color[rgb]{0,0,0}$wz^2$}%
}}}}
\put(-5399,-1336){\makebox(0,0)[b]{\smash{{\SetFigFontNFSS{12}{14.4}{\rmdefault}{\mddefault}{\updefault}{\color[rgb]{0,0,0}$zw$}%
}}}}
\put(-6599,-1336){\makebox(0,0)[rb]{\smash{{\SetFigFontNFSS{12}{14.4}{\rmdefault}{\mddefault}{\updefault}{\color[rgb]{0,0,0}$w$}%
}}}}
\put(-2399,1439){\makebox(0,0)[b]{\smash{{\SetFigFontNFSS{12}{14.4}{\rmdefault}{\mddefault}{\updefault}{\color[rgb]{0,0,0}b)}%
}}}}
\put(-7199,1439){\makebox(0,0)[b]{\smash{{\SetFigFontNFSS{12}{14.4}{\rmdefault}{\mddefault}{\updefault}{\color[rgb]{0,0,0}a)}%
}}}}
\put(-599,-361){\makebox(0,0)[b]{\smash{{\SetFigFontNFSS{12}{14.4}{\rmdefault}{\mddefault}{\updefault}{\color[rgb]{0,0,0}$\alpha$}%
}}}}
\end{picture}%

%% file: dm-newton.pdf_t
\begin{picture}(0,0)%
\includegraphics{dm-newton.pdf}%
\end{picture}%
\setlength{\unitlength}{2368sp}%
\begingroup\makeatletter\ifx\SetFigFontNFSS\undefined%
\gdef\SetFigFontNFSS#1#2#3#4#5{%
  \reset@font\fontsize{#1}{#2pt}%
  \fontfamily{#3}\fontseries{#4}\fontshape{#5}%
  \selectfont}%
\fi\endgroup%
\begin{picture}(3636,4377)(-7814,-2632)
\put(-5399,1514){\makebox(0,0)[b]{\smash{{\SetFigFontNFSS{12}{14.4}{\rmdefault}{\mddefault}{\updefault}{\color[rgb]{0,0,0}$z^2w^3$}%
}}}}
\put(-7799,389){\makebox(0,0)[b]{\smash{{\SetFigFontNFSS{12}{14.4}{\rmdefault}{\mddefault}{\updefault}{\color[rgb]{0,0,0}$bw^2$}%
}}}}
\put(-5399,389){\makebox(0,0)[b]{\smash{{\SetFigFontNFSS{12}{14.4}{\rmdefault}{\mddefault}{\updefault}{\color[rgb]{0,0,0}$z^2w^2$}%
}}}}
\put(-6599,389){\makebox(0,0)[b]{\smash{{\SetFigFontNFSS{12}{14.4}{\rmdefault}{\mddefault}{\updefault}{\color[rgb]{0,0,0}$zw^2$}%
}}}}
\put(-5399,-2536){\makebox(0,0)[b]{\smash{{\SetFigFontNFSS{12}{14.4}{\rmdefault}{\mddefault}{\updefault}{\color[rgb]{0,0,0}$a^3z^2$}%
}}}}
\put(-4199,389){\makebox(0,0)[lb]{\smash{{\SetFigFontNFSS{12}{14.4}{\rmdefault}{\mddefault}{\updefault}{\color[rgb]{0,0,0}$w^2z^3$}%
}}}}
\put(-4199,-1336){\makebox(0,0)[lb]{\smash{{\SetFigFontNFSS{12}{14.4}{\rmdefault}{\mddefault}{\updefault}{\color[rgb]{0,0,0}$a^2wz^3$}%
}}}}
\put(-5399,-1336){\makebox(0,0)[b]{\smash{{\SetFigFontNFSS{12}{14.4}{\rmdefault}{\mddefault}{\updefault}{\color[rgb]{0,0,0}$z^2w$}%
}}}}
\put(-6599,-1336){\makebox(0,0)[rb]{\smash{{\SetFigFontNFSS{12}{14.4}{\rmdefault}{\mddefault}{\updefault}{\color[rgb]{0,0,0}$zw$}%
}}}}
\end{picture}%